%% file: main.tex
\documentclass[sigconf, screen]{acmart}

\setcopyright{acmcopyright}
\acmPrice{15.00}
\acmDOI{10.1145/3468264.3468586}
\acmYear{2021}
\copyrightyear{2021}
\acmSubmissionID{fse21main-p453-p}
\acmISBN{978-1-4503-8562-6/21/08}
\acmConference[ESEC/FSE '21]{Proceedings of the 29th ACM Joint European Software Engineering Conference and Symposium on the Foundations of Software Engineering}{August 23--28, 2021}{Athens, Greece}
\acmBooktitle{Proceedings of the 29th ACM Joint European Software Engineering Conference and Symposium on the Foundations of Software Engineering (ESEC/FSE '21), August 23--28, 2021, Athens, Greece}

\AtBeginDocument{%
  \providecommand\BibTeX{{%
    \normalfont B\kern-0.5em{\scshape i\kern-0.25em b}\kern-0.8em\TeX}}}

\usepackage{caption}
\usepackage{subcaption}
\usepackage{listings, xcolor}
\usepackage{graphicx}   
\usepackage{amsmath}
\usepackage{microtype}
\input{model}

\begin{document}

\title{GLIB: Towards Automated Test Oracle for Graphically-Rich Applications}

\author{Ke Chen}
    \authornote{The first two authors contributed equally to this research.}
    \affiliation{%
      \institution{Fuxi AI Lab in Netease} 
      \city{Hangzhou}
    \country{China}}
    \email{chenke3@corp.netease.com}

\author{Yufei Li}
    \authornotemark[1]
\affiliation{%
  \institution{University of Texas at Dallas}
  \city{Dallas}
  \country{USA}}
  \email{yxl190090@utdallas.edu}

\author{Yingfeng Chen}
  \affiliation{%
    \institution{Fuxi AI Lab in Netease}
    \city{Hangzhou}
    \country{China}}
    \email{chenyingfeng1@corp.netease.com}

\author{Changjie Fan}
 \affiliation{%
  \institution{Fuxi AI Lab in Netease}
  \city{Hangzhou}
  \country{China}}
 \email{fanchangjie@corp.netease.com}

\author{Zhipeng Hu}
 \affiliation{%
 \institution{Fuxi AI Lab in Netease}
 \city{Hangzhou}
 \country{China}}
 \email{zphu@corp.netease.com}

\author{Wei Yang}
 \affiliation{%
  \institution{University of Texas at Dallas}
  \city{Dallas}
  \country{USA}}
 \email{wei.yang@utdallas.edu}

\renewcommand{\shortauthors}{Ke Chen, Yufei Li, Yingfeng Chen, Changjie Fan, Zhipeng Hu, and Wei Yang}

\input{Abstract}

\maketitle

\input{Introduction}

\input{Background}

\input{Study}

\input{Approach}

\input{Evaluation}

\input{Results_and_Analysis}

\input{Case_study}

\input{Application}

\input{Related}

\input{Threat}

\input{Conclusion}

\begin{acks}
We would like to thank Lei Ma, Zihe Song and Simin Chen for their help. We also thank the anonymous reviewers for their helpful feedback. This work was supported in part by UT Dallas startup funding \#37030034.
\end{acks}

\balance
\bibliographystyle{ACM-Reference-Format}
\bibliography{bibliography}

\appendix

\end{document}

%% file: model.tex
\def\UIissue{\textit{abnormal color block, random pixel noise, partial repetition, frame overlay, model missing, abnormal text, overexposed} and \textit{black border}\xspace}

\def\IssueNum{8\xspace}

\def\Compatibility{\textit{UI glitch, install failed, start failed, crash, app freeze, UI lags, black screen, network error} and \textit{other problems}\xspace}

\usepackage{xspace}  
\usepackage{enumitem}

\newcommand{\eg}{{\it e.g.,}\xspace}

\newcommand{\ie}{{\it i.e.,}\xspace}

%% file: Abstract.tex
\begin{abstract}

Graphically-rich applications such as games are ubiquitous with attractive visual effects of Graphical User Interface (GUI) that offers a bridge between software applications and end-users. However, various types of graphical glitches may arise from such GUI complexity and have become one of the main component of software compatibility issues. Our study on bug reports from game development teams in NetEase Inc. indicates that graphical glitches frequently occur during the GUI rendering and severely degrade the quality of graphically-rich applications such as video games. 
Existing automated testing techniques for such applications focus mainly on generating various GUI test sequences and check whether the test sequences can cause crashes. 
These techniques require constant human attention to captures non-crashing bugs such as bugs causing graphical glitches.
In this paper, we present the first step in automating the test oracle for detecting non-crashing bugs in graphically-rich applications. 
Specifically, we propose \texttt{GLIB} based on a code-based data augmentation technique to detect game GUI glitches.
We perform an evaluation of \texttt{GLIB} on 20 real-world game apps (with bug reports available) and the result shows that \texttt{GLIB} can achieve 100\% precision and 99.5\% recall in detecting non-crashing bugs such as game GUI glitches. Practical application of \texttt{GLIB} on another 14 real-world games (without bug reports) further demonstrates that \texttt{GLIB} can effectively uncover GUI glitches, with 48 of 53 bugs reported by \texttt{GLIB} having been confirmed and fixed so far.     
\end{abstract}

\ccsdesc[500]{Software and its engineering~Software testing and debugging}
\ccsdesc[500]{Computing methodologies~Neural networks}

\keywords{
Automated Test Oracle, Game Testing, GUI Testing, Deep Learning
}

%% file: Introduction.tex
\section{Introduction}

Graphically-rich applications (also short for apps) have been popular on mobile and personal computer (PC) platforms. With a growing number of complex visual effects such as advanced rendering, light and shadows, animation, and intensive media embedding being used to enhance the quality of GUI (also short for UI)~\cite{liu2020owl}, various graphical glitches may occur in the apps and severely impact user experience. Existing automatic UI testing techniques~\cite{zheng2019wuji} detect bugs by generating test sequences and check whether some crashes are caused. Therefore, these techniques require constant human attention to capture the UI glitch-inducing bugs. However, there are quantities of UI glitches that can severely degrade graphically-rich apps’ usability but not induce crashes in practical scenarios. Hence, in this paper, we make a first step in addressing the lack of oracle problem for graphically-rich apps. Specifically, we propose an automated test oracle for detecting UI glitches in game apps.

\input{Others/figure1}

Recent image-based UI testing techniques~\cite{liu2020owl,zhao2020seenomaly} demonstrate that adding images with versatile UI display issues to the training datasets can help improve the performance of Convolutional Neural Network (CNN)-based detection models in non-game mobile apps. For example, \texttt{Owl Eyes}~\cite{liu2020owl} designs a heuristic-based data augmentation approach for generating abnormal screenshots on Rico dataset~\cite{deka2017rico} by mimicking the symptom of real-world UI display issues. Its main methodology is to classify UI display issues into five classes and design each issue generation rule according to its features. With a large amount of generated UI screenshots, \texttt{Owl Eyes} improves the effectiveness of detecting UI glitches in non-game apps significantly.

However, we observe that the existing heuristic-based data augmentation approach applied in \texttt{Owl Eyes} cannot accurately reflect the UI glitch issues in graphically-rich applications, especially in game scenarios due to three main reasons. First, their manually-defined rules require human inspection on screenshots of UI glitches and humans may miss certain unnoticeable but important patterns.
Moreover, their generation process is to mimic the screenshots of UI glitches, thus the generated images may be infeasible to be generated by the real bugs in the program code. This approximation may cause false positives in the detection process.
Last, \texttt{Owl Eyes} mainly focuses on text-related UI display issues, whereas in game scenario, the UI display issues are typically text-irrelevant graphical glitches which may not be generalized by the heuristic rules defined by \texttt{Owl Eyes}.

To address these issues, we propose \texttt{GLIB}, an automated test oracle to detect UI glitch-related bugs. 
To enable better performance of \texttt{GLIB}, we develop a code-based data augmentation approach to augment the training data for \texttt{GLIB} by injecting the buggy code snippets to the game apps and record the manifestation of the bugs (i.e., UI glitches). In this way, our generated screenshots contain real UI glitches so that the DL model can be trained with more precise datasets and potentially learn subtle patterns that humans may not observe. Moreover, our study of bugs' root causes can guide developers to debug with some empirical knowledge after detecting the UI glitches. 
Because some UI glitch issues occur in only parts of the UI screen area and human inspectors may miss the issues, we develop a technique based on the saliency map~\cite{simonyan2013deep} to localize the glitch regions with different bug categories so that the developers can easily determine whether and where our detected images have UI glitches.

To better evaluate the effectiveness of \texttt{GLIB}, we create a testing dataset consisting  app screenshots with and without UI glitches from real-world game bug reports. Evaluation on the testing dataset demonstrates that \texttt{GLIB} can achieve 0.9\% and 76.7\% boost in precision and recall compared to the prediction results of the model trained without data augmentation, leading to 100\% precision, 99.5\% recall, and 99.8\% F-1. Moreover, we evaluate the practical usefulness of \texttt{GLIB} by detecting UI glitches in 14 real-world games with different platforms and engines, the practical application result shows that our model can successfully spot previously undetected UI glitch issues and help developers to fix the bug.

The contributions of this article are as follows:
\begin{itemize}
    \item Our work\footnote{Code to reproduce our experiments is available at \url{https://github.com/GLIB-game/GLIB.git}}~\cite{GLIB} is the first to systematically investigate UI glitch issues in real-world graphically-rich apps. We create a large-scale dataset of screenshots with UI glitches and release the data~\cite{GLIB_dataset} for follow-up studies.
    
    \item Based on our characteristic study on the root causes of graphical UI glitches, we propose a code-based training data augmentation approach that can be applied in real-world game apps to generate UI glitches. Our study can also guide developers to find and fix the bug after detecting game UI glitches.
    
    \item We propose a CNN-based model for detecting images with UI display issues, and leverage saliency map to localize the glitch region in the UI.

\end{itemize}

%% file: Others/figure1.tex
\begin{figure}[h]
     \centering
     \begin{subfigure}[b]{0.23\textwidth}
     \includegraphics[height=2.2cm,width=\textwidth]{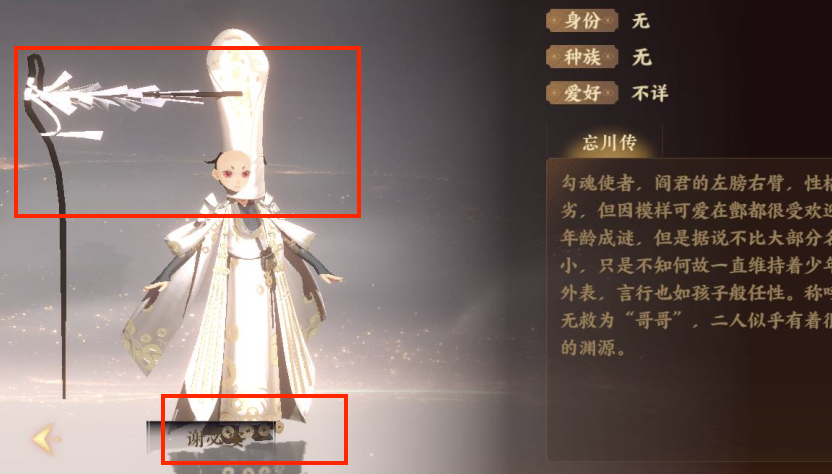}
         \label{fig:exp1}
     \end{subfigure}
     \hfill
     \begin{subfigure}[b]{0.23\textwidth}
     \includegraphics[height=2.2cm,width=\textwidth]{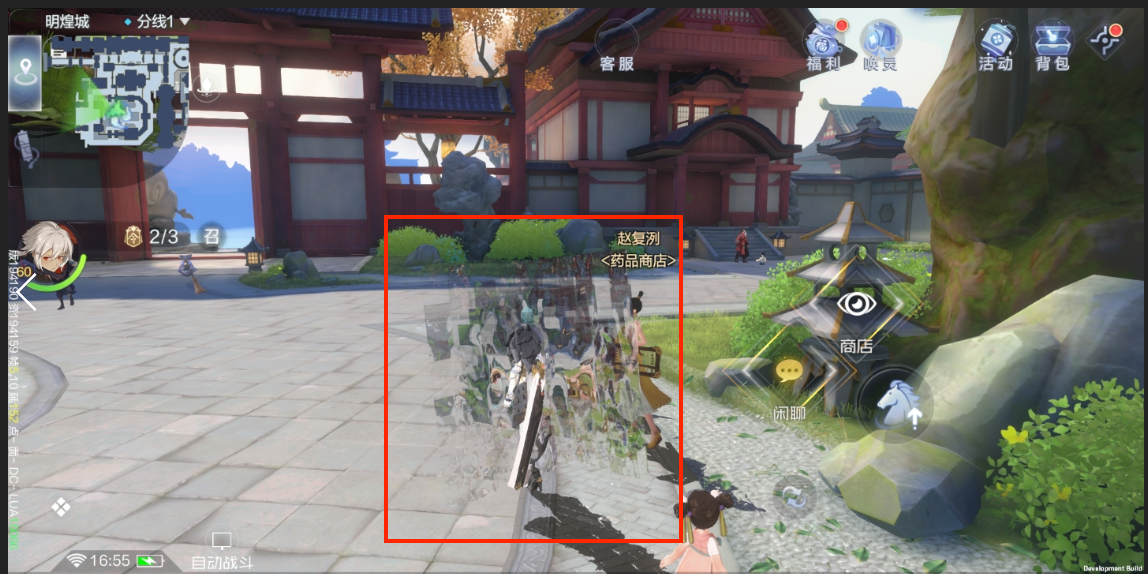}
         \label{fig:exp2}
     \end{subfigure}
    \caption{Examples of game UI glitches.}
    \label{fig:show_exp}
\end{figure}

%% file: Background.tex
\section{Background}

Game testing company TestBird~\cite{TestBird} collected and tested 11,476 mobile game apps in 2020 and reported 552,851 relevant compatibility issues. According to the statistics of 300 terminals tested for each game, the average number of game compatibility issues is 44 and the average pass rate is 86.41\%. TestBird analyzed all these compatibility issues and classified them into 10 categories, namely, \Compatibility. Among them, UI glitch and game crash are the two largest categories with UI glitch accounting for 39.95\% and game crash occupying 28.76\% of compatibility issues. The detailed compatibility issue distribution is shown in Figure~\ref{fig:compatibility}. Particularly, UI glitch occurs and has been the most severe compatibility issue in nearly every tested mobile game. TestBird also investigated the proportion of game engines on mobile game applications as shown in Figure~\ref{fig:engine}, among which Unity3d (also short for Unity) is the most prevalent one and hence our following study is based on the Unity game engine. Another game company WeTest~\cite{WeTest} tested all Tencent~\cite{TecentGame} mobile games and summarized the compatibility issues into eight categories among which UI glitch is also the largest issue and accounts for 47.5\% of all the problems. Specifically, the percentage of UI glitches increased by 11\% compared to last year and among them the proportion of problems such as \textit{abnormal color block} and \textit{random noise} has increased.

\input{Others/figure2}

\input{Others/figure3}

%% file: Others/figure2.tex
\begin{figure}[h]
     \centering
     \begin{subfigure}[b]{0.225\textwidth}
         \centering
     \includegraphics[width=\textwidth]{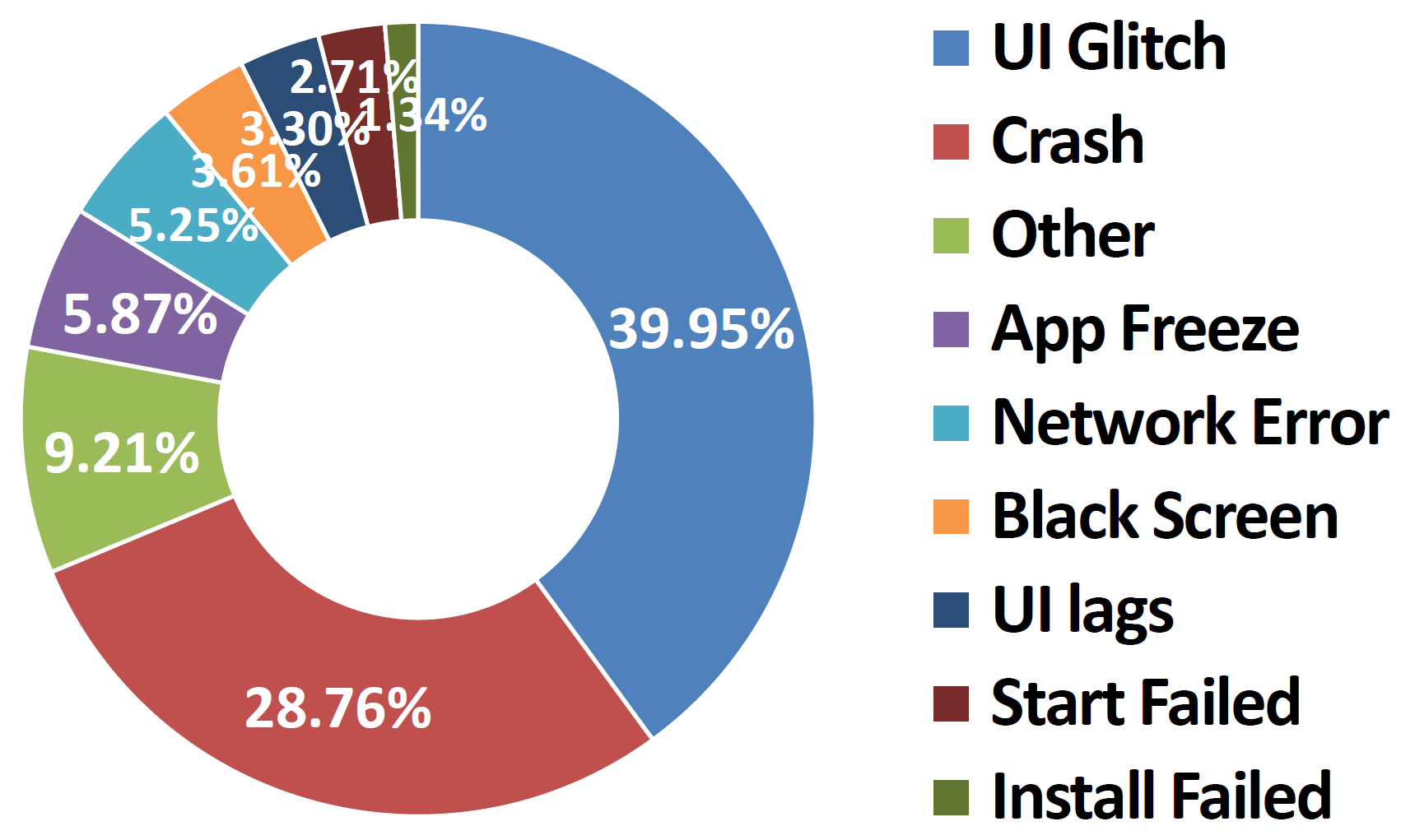}
        \caption{Game issue proportion} \label{fig:compatibility}
         
     \end{subfigure}
     \hfill
     \begin{subfigure}[b]{0.245\textwidth}
         \centering
     \includegraphics[width=\textwidth]{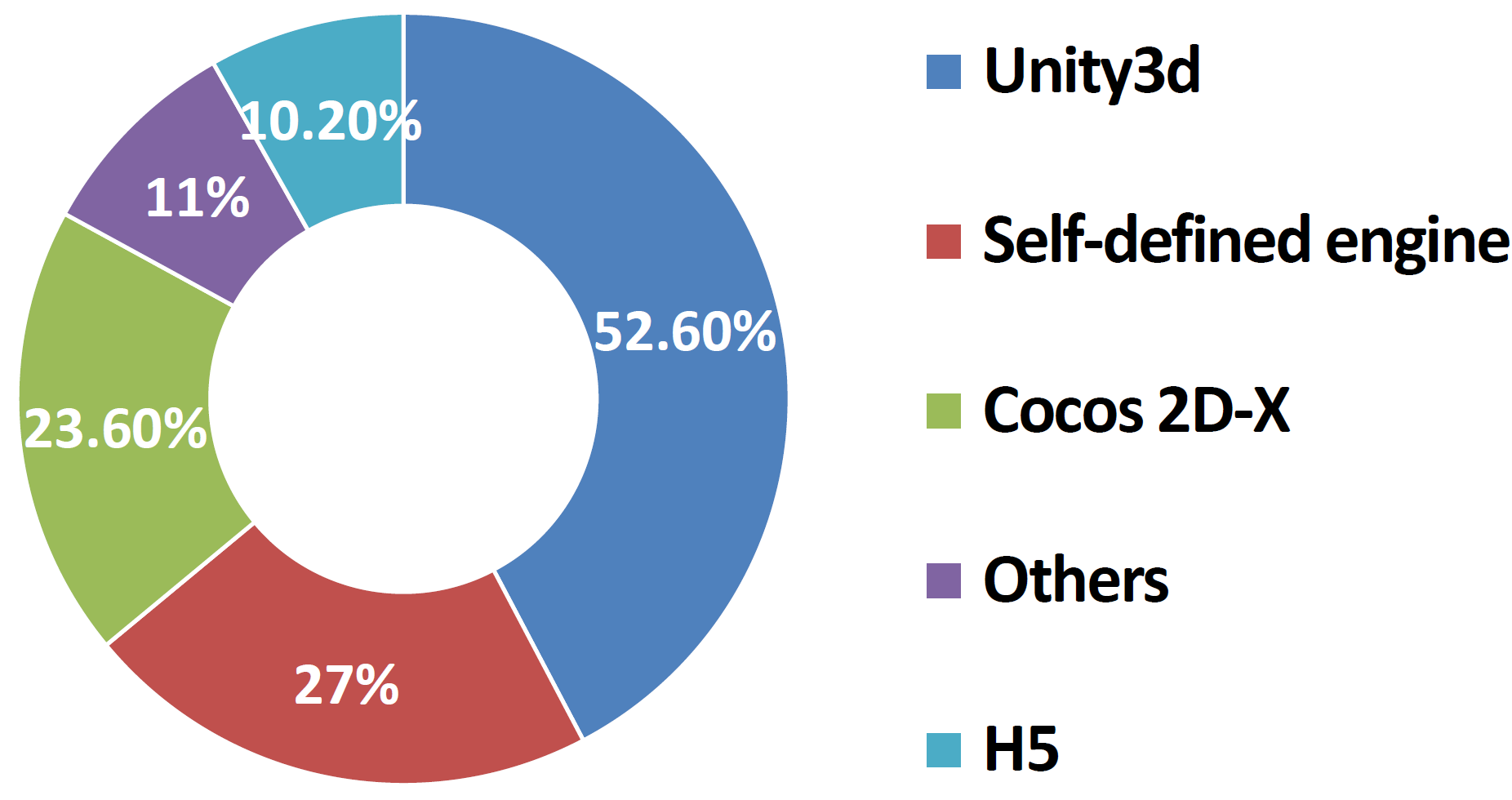}
         \caption{Game engine proportion}
         \label{fig:engine}
     \end{subfigure}

        \caption{Test report from TestBird.}
        \label{fig:test_report}
\end{figure}

%% file: Others/figure3.tex
\begin{figure*}
     \centering
     \begin{subfigure}[b]{0.24\textwidth}
         \centering
         \includegraphics[height=2.1cm, width=\textwidth]{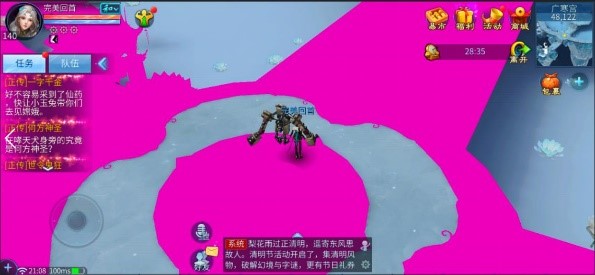}
         \caption{Abnormal color block}
         \label{fig:Abnormal color block}
     \end{subfigure}
     \hfill
     \begin{subfigure}[b]{0.24\textwidth}
         \centering
         \includegraphics[height=2.1cm,width=\textwidth]{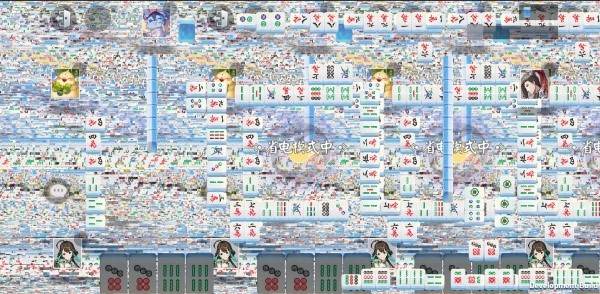}
         \caption{Random noise}
         \label{fig:Random noise}
     \end{subfigure}
     \hfill
     \begin{subfigure}[b]{0.24\textwidth}
         \centering
         \includegraphics[height=2.1cm,width=\textwidth]{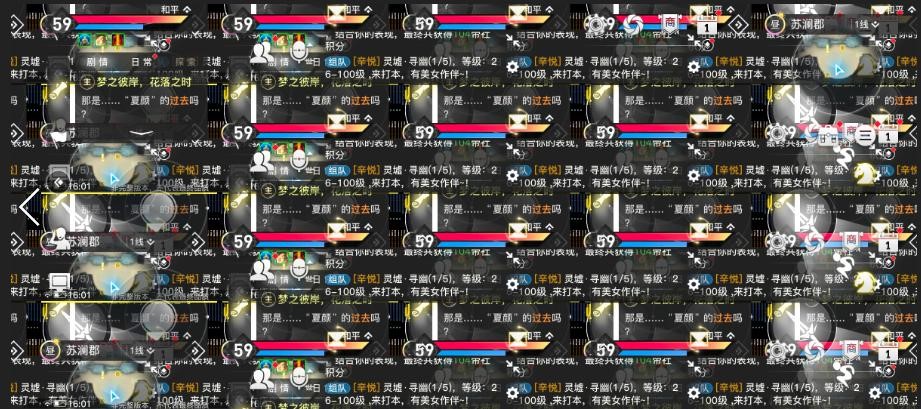}
         \caption{Partial repetition}
         \label{fig:Partial repetition}
     \end{subfigure}
     \hfill
     \begin{subfigure}[b]{0.24\textwidth}
         \centering
         \includegraphics[height=2.1cm,width=\textwidth]{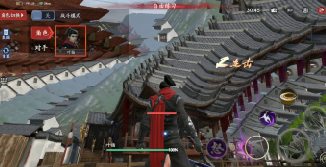}
         \caption{Frame overlay}
         \label{fig:Frame overlay}
     \end{subfigure}
     \hfill
     \begin{subfigure}[b]{0.24\textwidth}
         \centering
         \includegraphics[height=2.1cm,width=\textwidth]{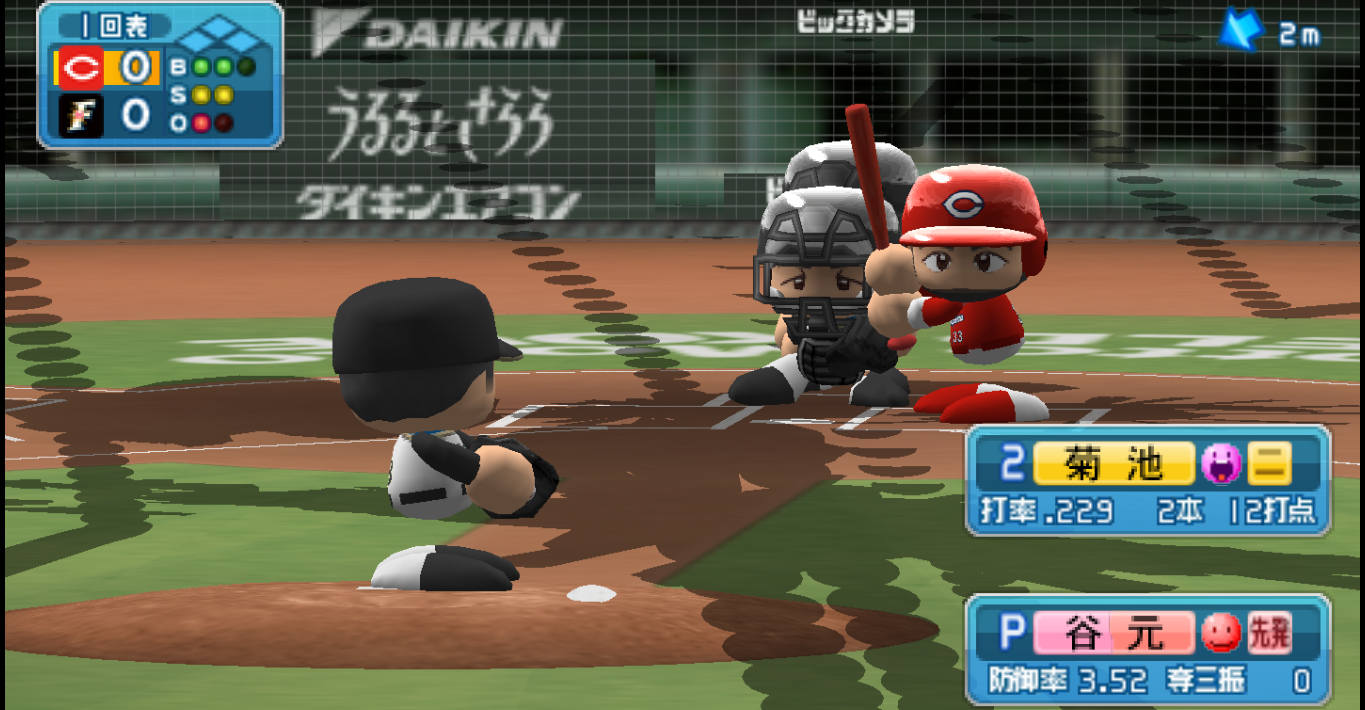}
         \caption{Object missing}
         \label{fig:Object missing}
     \end{subfigure}
     \hfill
     \begin{subfigure}[b]{0.24\textwidth}
         \centering
         \includegraphics[height=2.1cm,width=\textwidth]{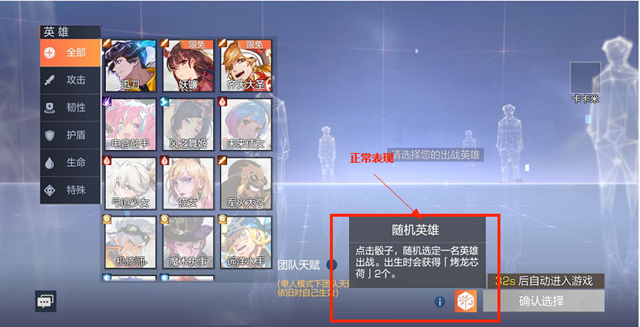}
         \caption{Abnormal text}
         \label{fig:Abnormal text}
     \end{subfigure}
     \hfill
     \begin{subfigure}[b]{0.24\textwidth}
         \centering
         \includegraphics[height=2.1cm,width=\textwidth]{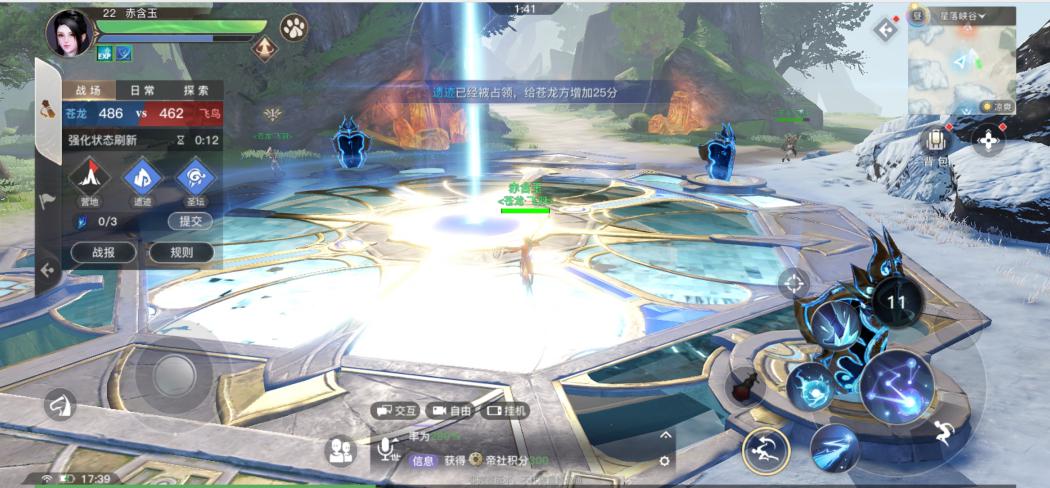}
         \caption{Overexposed}
         \label{fig:Highlight}
     \end{subfigure}
     \hfill
     \begin{subfigure}[b]{0.24\textwidth}
         \centering
         \includegraphics[height=2.1cm,width=\textwidth]{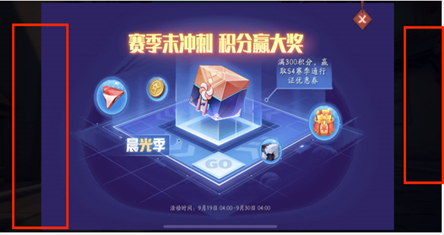}
         \caption{Black border}
         \label{fig:Black border}
     \end{subfigure}
     
        \caption{Examples of eight categories of game UI glitches.}
        \label{fig:UI glitch}
\end{figure*}

%% file: Study.tex
\section{Characteristic Study}

\definecolor{codegreen}{rgb}{0,0.6,0}
\definecolor{codegray}{rgb}{0.5,0.5,0.5}
\definecolor{codepurple}{rgb}{0.58,0,0.82}
\definecolor{backcolour}{rgb}{1,1,1}

\lstdefinestyle{mystyle}{
  backgroundcolor=\color{backcolour},   commentstyle=\color{codegreen},
  keywordstyle=\color{magenta},
  numberstyle=\tiny\color{codegray},
  stringstyle=\color{codepurple},
  basicstyle=\ttfamily\footnotesize,
  breakatwhitespace=false,         
  breaklines=true,                 
  captionpos=b,                    
  keepspaces=true,                 
  numbers=left,                    
  numbersep=5pt,                  
  showspaces=false,                
  showstringspaces=false,
  showtabs=false,                  
  tabsize=2,
  escapeinside={\%*}{*)}
}

\definecolor{codelightgreen}{rgb}{0.929,1,0.902}

\lstdefinestyle{mystyle2}{
  backgroundcolor=\color{codelightgreen},
  keywordstyle=\color{magenta},
  numberstyle=\tiny\color{codegray},
  stringstyle=\color{codepurple},
  basicstyle=\ttfamily\footnotesize,
  breakatwhitespace=false,         
  breaklines=true,                 
  captionpos=b,                    
  keepspaces=true,                 
  numbersep=5pt,                  
  showspaces=false,                
  showstringspaces=false,
  showtabs=false,                  
  tabsize=2
}

\lstdefinestyle{my-java-style}{
    backgroundcolor=\color{backcolour},  
    language=my-java,
    basicstyle=\footnotesize,   
    captionpos=b,                  
    alsoletter={.},         
    escapeinside={\%*}{*)},       
    extendedchars=true,          
    frame=tb,                     
    keepspaces=true,   
    keywordstyle=\color{javapurple}\bfseries,  
    rulecolor=\color{black},
    tabsize=4
}

Before we build a model to detect game UI glitches, we collected quantities of UI glitch issues that appeared in the real-world game apps. Our study aims to answer the following two questions:

\begin{itemize}[leftmargin=*]

\item\textbf{RQ1}: What is the general manifestation of UI graphical glitches in mobile game apps?
    
\item\textbf{RQ2}: What are the bug causes of these Game UI glitches?

\end{itemize}

\subsection{Data Collection}

To better understand the UI glitch issues in real-world mobile game apps, we collect the 466 bug reports of 20 NetEase~\cite{NetEase} Android games belonging to different categories such as Adventure Game, Action Game, First-Person Shooter Game, Role Playing Game, etc. with 2,418 UI glitch images. The main reason we focus on mobile game UI display issues is that compatibility issues between software and hardware frequently appear on mobile devices. Take Android as an example, nowadays more than 10 major versions of the Android operating system (OS) run on 24,000+ distinct device models with different screen resolutions~\cite{wei2016taming}. Because most of the abnormal images are photos captured from the external camera with some annotation rather than the screenshots, we preprocess these images by remaining the screenshots and excluding photos and images with system-related bugs. Finally, we obtain 201 filtered graphical glitch screenshots and use them for our characteristic study.

\subsection{Manifestation of Game UI Glitches (RQ1)}
\label{subsec:manifestation}

Given those collected screenshots, we found that UI glitches are actually versatile in terms of their manifestation and that different types of UI glitches may appear with different frequencies thus has a different level of impact on the game usability. Therefore, categorization of these issues would facilitate our study, design, and evaluation of the related approach.  Following the Card Sorting~\cite{spencer2009card} and adapt the technique to the game scenario, we categorize these bugs into \IssueNum categories, namely, \UIissue, and statistic their proportion of occurrence with details as follows:

{\textbf{Abnormal color block (56\%).}} As shown in Figure~\ref{fig:Abnormal color block}, abnormal color blocks stretch and cover the UI graph. The main cause is that some material is missing or the camera responsible for rendering pixel RGB values is incorrectly turned off.

{\textbf{Random noise (17\%).}} As shown in Figure~\ref{fig:Random noise}, quantities of color pixels randomly distribute over the whole screen or specific area. The main cause is that the camera is incorrectly turned off.

{\textbf{Partial repetition (12\%).}} As shown in Figure~\ref{fig:Partial repetition}, part of the UI area is repeated or mirrored. Disabling camera or post-processing error (GPU does not support the rendering effect or incorrect render logic) may result in this glitch issue.

{\textbf{Frame overlay (6\%).}} As shown in Figure~\ref{fig:Frame overlay}, the frame in the previous time step overlaps the current frame. The wrong value of the camera's clearflag might be the main reason.

{\textbf{Object missing (3\%).}} As shown in Figure~\ref{fig:Object missing}, the UI model lacks part of its component. This may be caused by the incorrect values of the alpha channel in the model texture.

{\textbf{Abnormal text (2\%).}} As shown in Figure~\ref{fig:Abnormal text}, multiple pieces of texts are located in the wrong area and may cover the characters or other objects. The main reason for this glitch issue is that the UI is not adapted to the screen resolution.

{\textbf{Overexposed (2\%).}} As shown in Figure~\ref{fig:Highlight}, the whole (or part of) UI scene is too bright or overexposed. The main reason is that the intermediate result is stored in a low precision variable and the result is clipped or overflows.

{\textbf{Black border (2\%).}} As shown in Figure~\ref{fig:Black border}, the UI image is not flattened to cover the whole screen and leaves the black borders on both sides because the display resolution (\eg 854 $\times$ 480) and aspect ratio of some special devices are not considered by the developer.

To ensure the completeness of our study, we asked several game testing experts from the company’s development teams to confirm that our summarized UI glitches cover all the common UI issues in their games including not only mobile apps but also other platforms such as PC and PS4. We also demonstrate that our \texttt{GLIB} can be applied to various types of games on distinct platforms and precisely detect UI bug issues in RQ4.

\subsection{Bug Causes of Game UI Glitches (RQ2)}
\label{subsec:root cause}

After we collect quantities of game UI glitch samples and have a common sense of UI display issues and their threat to the user's game experience, a more important thing is to understand the bug causes of these glitch issues. To common sense, the reason for game UI display issues might be the defects of hardware (\eg GPU-related issues) or the wrong setting of rendering special effects. To facilitate the visual understanding in detecting UI display issues, we focus on explaining the root causes in terms of source-code level (bugs). By doing so, we collect the historical commit diff of various game apps and categorize the bug issues into 4 major types. For each bug fix example, the code marked with green color is the missing part of the original bug code. 

{\textbf{Rendering cameras are turned off incorrectly.}} Cameras in the Unity engine are the devices that capture the color and depth information of the game world and display the whole scene to the players. A game scene can hold an unlimited number of cameras, with different objects probably rendered by different cameras. If one camera is turned off unexpectedly, the render results may be substituted by any memory block which has not been initialized, thus the RGB values of the corresponding objects in the image can be randomized, and the manifestation of these random pixel values in game UI display issues will most likely be \textit{abnormal color block}, \textit{random noise} or \textit{object missing}. The bug fix procedure of camera enabled error is shown in Listing 1.

\lstdefinestyle{sharpc}{language=[Sharp]C, rulecolor=\color{blue!80!black}}
\lstset{style=sharpc}
\lstset{style=mystyle}

\begin{lstlisting}[style=mystyle, xleftmargin=2.5ex]
	targetCamera.targetTexture = originRT;
\end{lstlisting}
\vspace{-6pt}
\lstset{style=sharpc}
\begin{lstlisting}[firstnumber=last, style=mystyle2, xleftmargin=2.5ex]
+	targetCamera.enabled = true;
\end{lstlisting}
\vspace{-6pt}
\begin{lstlisting}[style=mystyle, , xleftmargin=2.5ex, firstnumber=last, caption=Bug fix procedure of camera enable error.]
	if (UIManager.inst != null && UIManager.inst.uiCamera != null)
	{
		UIManager.inst.uiCamera.enabled = true;
	}
\end{lstlisting}

\input{Others/figure4}

{\textbf{Wrong settings of camera's clearflag.}} Cameras in the Unity engine typically clear the color and depth information on the screen before rendering image frames, and the clearflag function of a camera determines how the color buffer and depth buffer are cleared. If the clearflag instruction is modified incorrectly, the depth and color settings of the scene may get chaotic, \eg if there is a cube moving randomly in the scene with the blue and gray parts as the background (\ie the depth is infinite), and the camera's clearflag is incorrectly set as "\textit{Don't Clear}", the color and depth buffers of the previous frame will remain and cause frame repeatedly appear at each time step. The white and red objects in Figure~\ref{fig:cube} are camera and cube to be rendered, respectively, and the rendering result is in Figure~\ref {fig:cube_move}. This bug is regarded as the main cause of \textit{frame overlay}. We fix this bug by setting the camera's clearflag according to the depth buffer. The bug fix procedure of camera clearflag error is shown in Listing 2.

\input{Others/figure5}
\input{Others/figure6}

\begin{lstlisting}[style=mystyle, xleftmargin=2.5ex]
	var originRT = targetCamera.targetTexture;
	targetCamera.targetTexture = rt;
	targetCamera.Render();
	CameraPostEffect.instance.doPostEffects(rt, postRT);
	if (UIManager.inst != null && UIManager.inst.uiCamera != null)
	{
		if (uiCameraOn)
		{
			UIManager.inst.uiCamera.Render();
			var originUIRT = UIManager.inst.uiCamera.targetTexture;
			UIManager.inst.uiCamera.targetTexture = postRT;
			UIManager.inst.uiCamera.Render();
\end{lstlisting}	
\vspace{-6pt}
\begin{lstlisting}[style=mystyle2, firstnumber=last, xleftmargin=2.5ex]
+			UIManager.inst.uiCamera.clearFlags = CameraClearFlags.Depth;
\end{lstlisting}
\vspace{-6pt}
\begin{lstlisting}[style=mystyle, , xleftmargin=2.5ex, firstnumber=last, caption=Bug fix procedure of camera clearflag error.]
			UIManager.inst.uiCamera.targetTexture = originUIRT;
		}
	}
\end{lstlisting}

{\textbf{Post-processing special effects of the previous scene are not cleared in time.}} Adding post-processing can apply various kinds of filters or effects to the camera’s image buffer before an image is displayed on the screen, and this post-processing technique drastically improves the visual expression of the scene. But if the post-processing effect is added incorrectly or if the effect is not cleared in time when the scene changes, the image content will become scrambled, even mess up the whole scene. For example, when one enters a scene without any post-process effects as in Figure~\ref{fig:normalscence}, it looks like a man is standing on the ground. However, if a game developer adds a mirror effect to the previous scene (\eg a lake scene in Figure~\ref{fig:mirroreffect}), and steps into the scene without clearing the post-process effect in time, the image then becomes symmetric as shown in Figure~\ref{fig:postprocess} as if there is a lake in the scene. This post-processing special effect bug is likely to cause \textit{partial repetition} issue and the bug fix code is shown in Listing 3.

\lstset{style=sharpc}
\lstset{style=mystyle}
\begin{lstlisting}[xleftmargin=2.5ex]
	private static LuaFunction m_hookOnDisable = null;
	private void OnDisable() {
	if (m_hookOnDisable != null) { if(GameBaseObject.Inst.InvokeNewHook(m_hookOnDisable, this)) return; }
		if (CameraPostEffect.instance == null || image.texture == null)
		{
			return;
		}
		CameraPostEffect.instance.ReleaseUIBloomTarget();
\end{lstlisting}
\vspace{-6pt}
\begin{lstlisting}[style=mystyle2, firstnumber=last, xleftmargin=2.5ex]
+		CameraPostEffect.instance.ClearPostRenderRT();
\end{lstlisting}
\vspace{-6pt}
\begin{lstlisting}[style=mystyle, , xleftmargin=2.5ex, firstnumber=last,caption=Bug fix procedure of incorrect camera post-processing effect error.]
	}
\end{lstlisting}

\input{Others/figure7}

{\textbf{GPU-related rendering bugs.}} Some UI glitches are caused by GPU driver bugs or GPU-related rendering bugs. For instance, the version of the operating system (\eg Android, iOS) on the mobile device might be too old and its handling palettes on the GPU cause UI display issues or the wrong GPU rendering settings like skipping some buffering effect for faster program running might cause \textit{object missing} issue. Wrong GPU rendering settings may also lead to resolution adaption problem and \textit{text out of position} issue.

\input{Others/figure8}

%% file: Others/figure4.tex
\begin{figure}[t]
     \centering
     \begin{subfigure}[b]{0.23\textwidth}
         \centering
     \includegraphics[width=\textwidth]{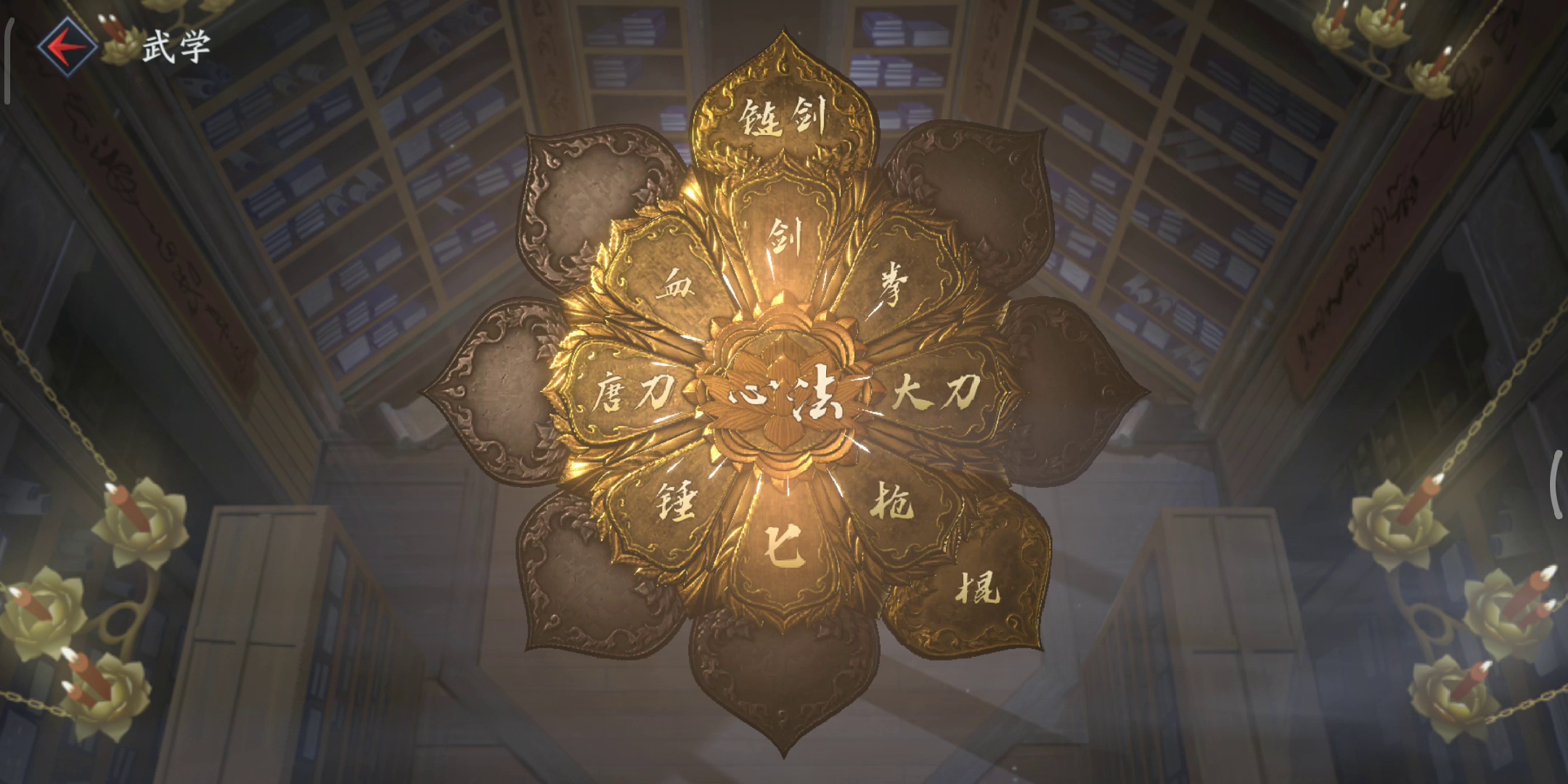}
         \label{fig:ce_normal}
     \end{subfigure}
     \hfill
     \begin{subfigure}[b]{0.23\textwidth}
         \centering
     \includegraphics[width=\textwidth]{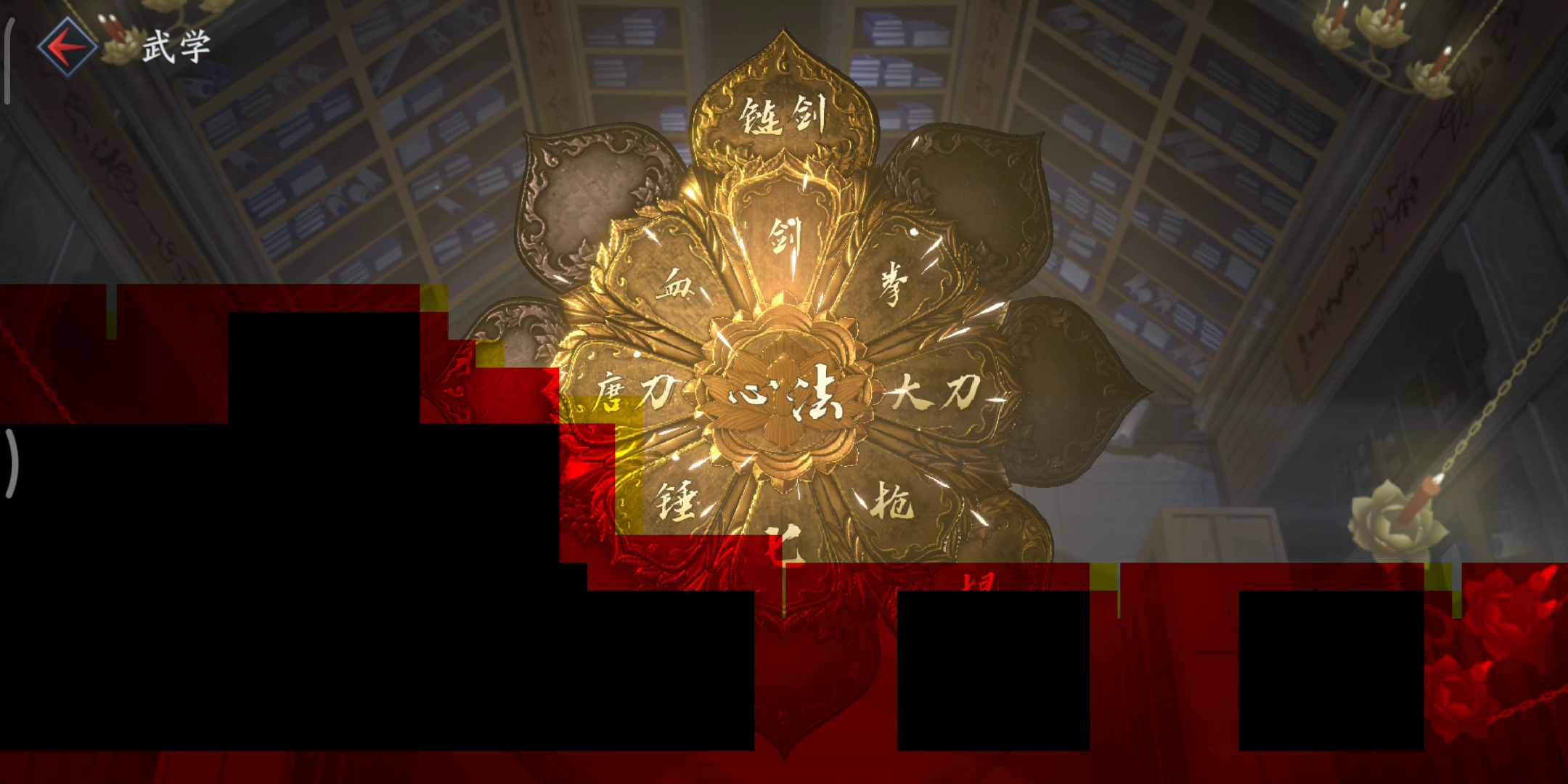}
         \label{fig:ce_error}
     \end{subfigure}
        \caption{Effect of incorrect camera turned off.}
        \label{fig:camera_enable_sample}
\end{figure}

%% file: Others/figure5.tex
 \begin{figure}[t]
     \centering
     \begin{subfigure}[b]{0.23\textwidth}
         \centering
     \includegraphics[width=\textwidth]{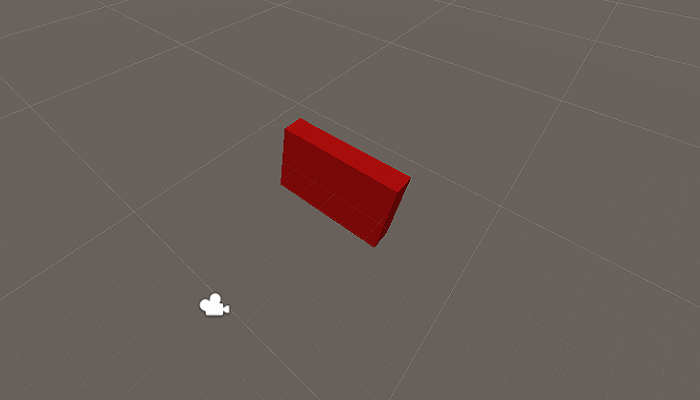}
         \caption{Rendering components}
         \label{fig:cube}
     \end{subfigure}
     \hfill
     \begin{subfigure}[b]{0.23\textwidth}
         \centering
     \includegraphics[width=\textwidth]{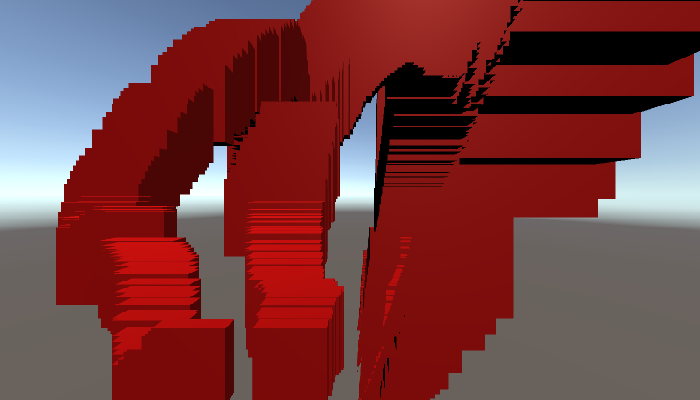}
         \caption{Rendering effect}
         \label{fig:cube_move}
     \end{subfigure}
        \caption{Frame overlay generation process.}
        \label{fig:clearflag}
\end{figure}

%% file: Others/figure6.tex
\begin{figure}[t]
     \centering
     \begin{subfigure}[b]{0.23\textwidth}
         \centering
     \includegraphics[width=\textwidth]{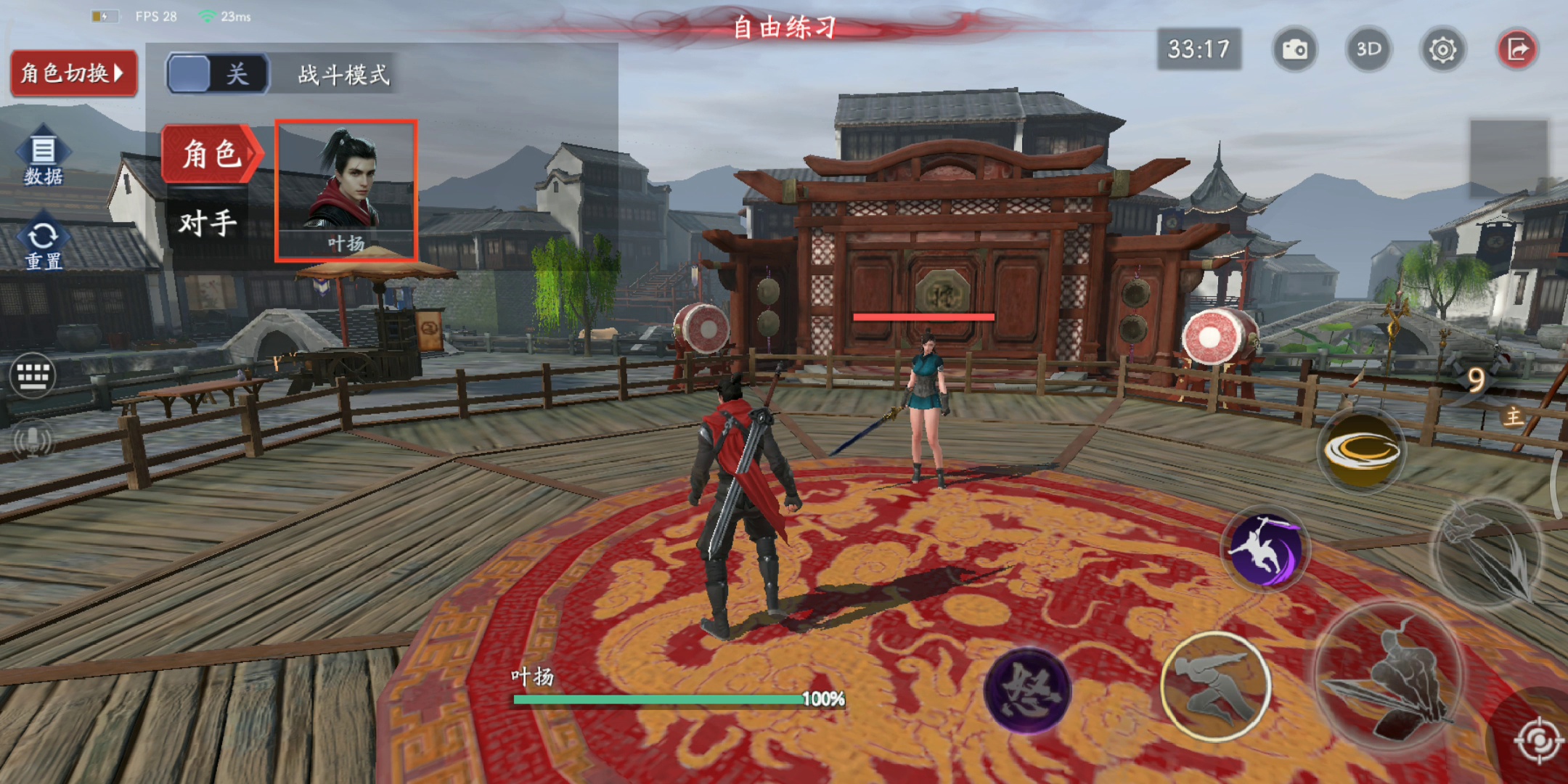}
         \label{fig:cf_normal}
     \end{subfigure}
     \hfill
     \begin{subfigure}[b]{0.23\textwidth}
         \centering
     \includegraphics[width=\textwidth]{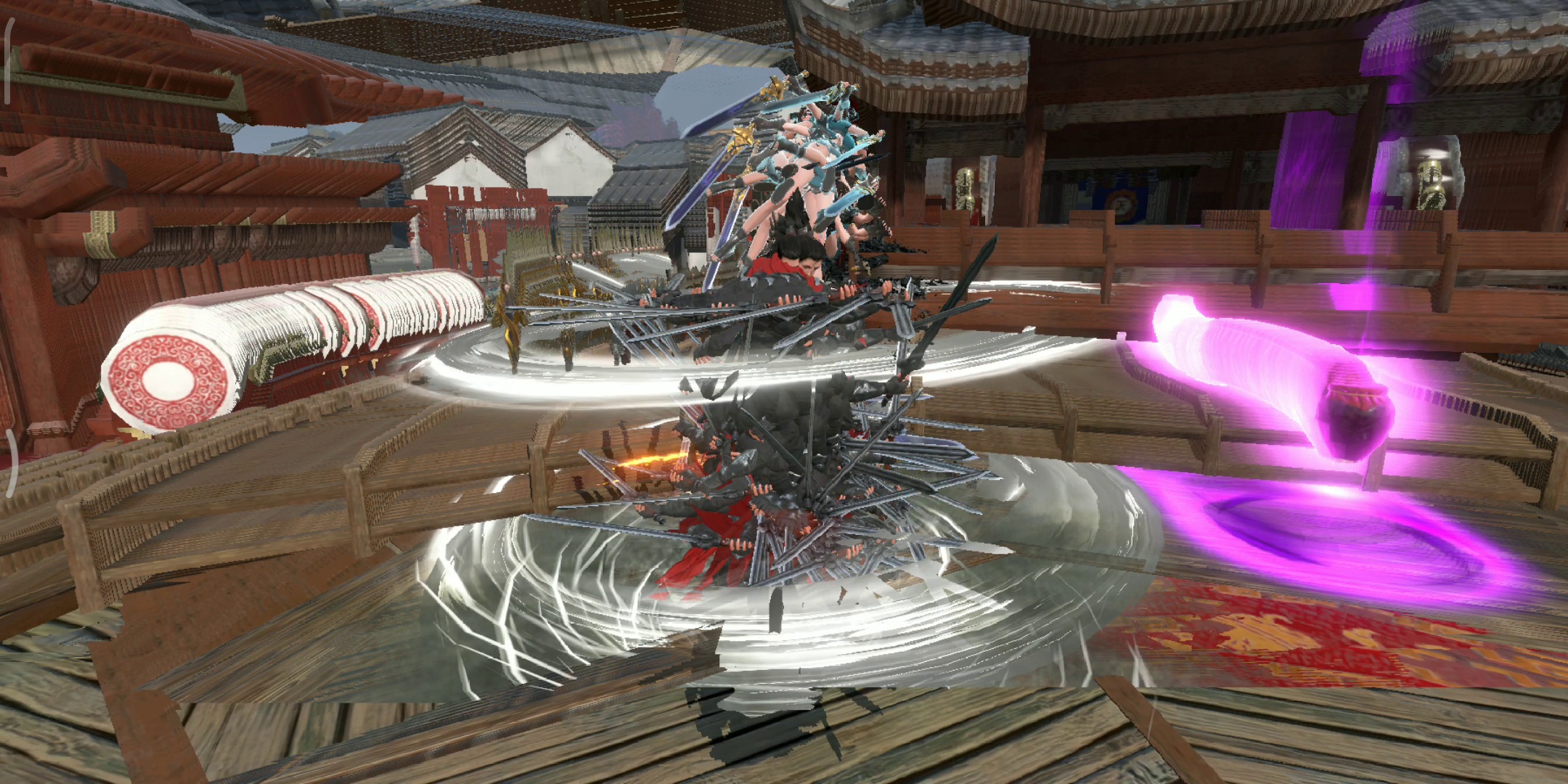}
         \label{fig:cf_error}
     \end{subfigure}
        \caption{Effect of wrong settings of camera clearflag.}
        \label{fig:camera_clearflag_sample}
\end{figure}

%% file: Others/figure7.tex
\begin{figure}[t]
     \centering
     \begin{subfigure}[h]{0.155\textwidth}
         \centering
     \includegraphics[height=1.5cm, width=\textwidth]{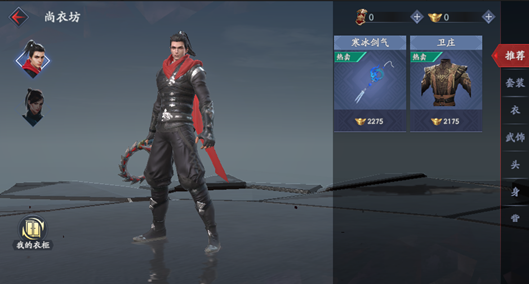}
         \caption{Raw scene}
         \label{fig:normalscence}
     \end{subfigure}
     \hfill
     \begin{subfigure}[h]{0.155\textwidth}
         \centering
     \includegraphics[height=1.5cm,width=\textwidth]{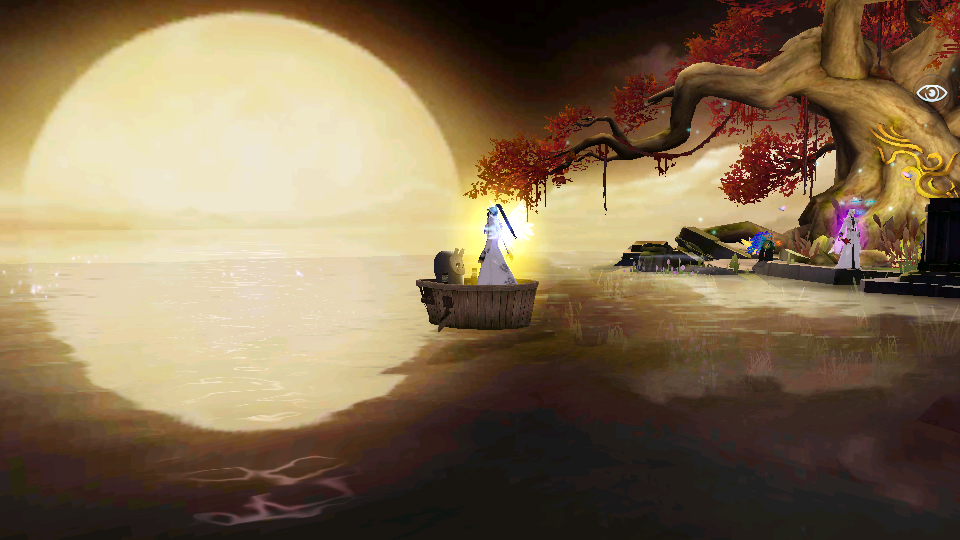}
         \caption{Mirror effect}
         \label{fig:mirroreffect}
     \end{subfigure}
     \hfill
     \begin{subfigure}[h]{0.155\textwidth}
         \centering
     \includegraphics[height=1.5cm,width=\textwidth]{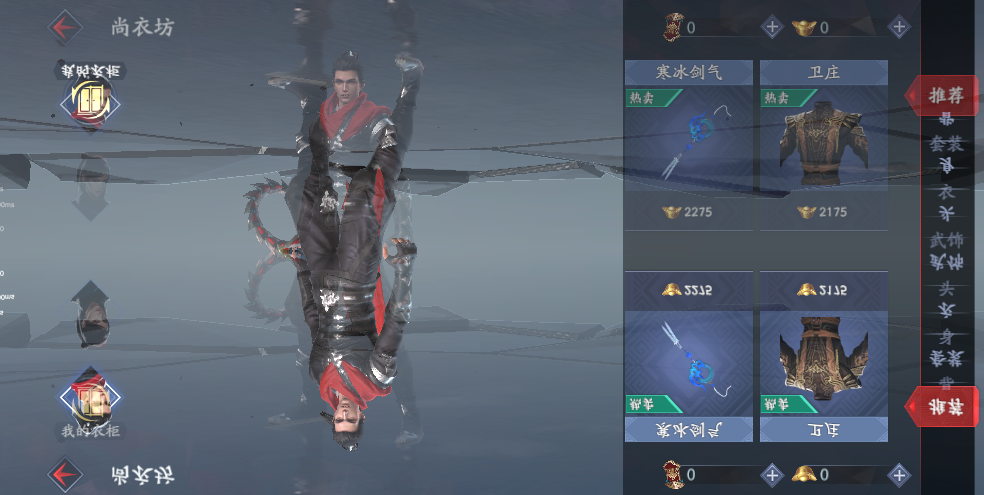}
         \caption{Synthetic scene}
         \label{fig:postprocess}
     \end{subfigure}
        \caption{Effect of post-processing error.}
        \label{fig:rc}
\end{figure}

%% file: Others/figure8.tex
\begin{figure*}
    \centering
    \includegraphics[width=\textwidth]{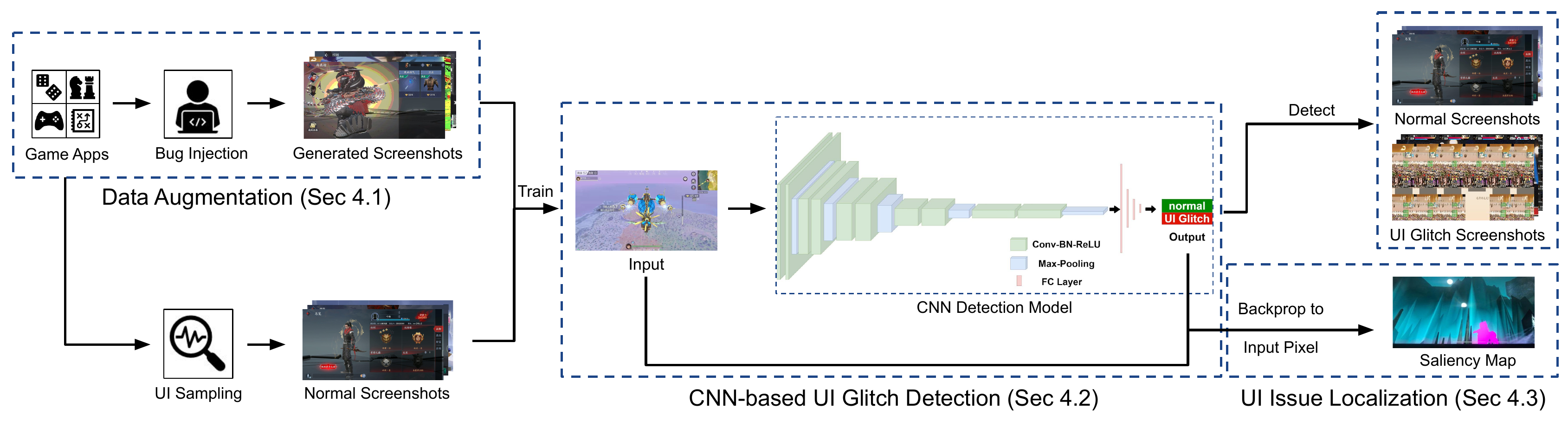}
    \caption{Overview of GLIB.}
    \label{fig:GLIB architecture}
\end{figure*}

%% file: Approach.tex
\section{Main Approach}

Our main approach \texttt{GLIB} consists of three parts, first, we propose a source-code based augmentation approach for generating quantities of abnormal game UI display images, then we design a CNN-based image recognition model to learn the pattern of various categories of game UI glitch issues and detect those screenshots with UI display issues, finally, we come up with a saliency map for automatic problem localization. Our \texttt{GLIB} frame is shown in Figure~\ref{fig:GLIB architecture}.

\subsection{Code-Based Data Augmentation}
\label{subsubsec:code_aug}

Training a powerful CNN model for visual recognition and UI issue detection requires quantities of data samples. For example, DenseNet~\cite{huang2019convolutional, huang2017densely} uses 50,000 samples from CIFAR~\cite{krizhevsky2009learning}, 73,257 images from SVHN~\cite{netzer2011reading} and 1.2 million images from ImageNet~\cite{deng2009imagenet} for training. Similarly, our proposed CNN model for UI glitch detection requires a large number of screenshots with versatile UI glitches. Nevertheless, our collected real-world game UI screenshots contain a small proportion of glitch images which also do not fully cover diverse categories of game UI glitches as we mentioned in Section~\ref{subsec:manifestation}. Therefore, we propose a code-based data augmentation approach based on the root causes we study in Section~\ref{subsec:root cause} for generating UI glitch problems by modifying the source code of various mobile game apps and making screenshots for typical scenes.

Particularly, when we inject the corresponding bug code into mobile game execution programs to force UI glitches to occur so that we can collect quantities of screenshots with UI display issues, we must ensure that only UI-related issues happen and other functions of the game apps are not affected (\eg do not crash) after their programs get updated. We wrap the bug code with execution parameter settings (which we also refer to as patch code) so that the execution program could get updated to the bug-injected version, and this patching technique is called \textit{hotfix}.

Our code-based UI glitch generation approach is automated and can be well-generalized to other games. With \textit{hotfix}, we only need to download the patch code that is pushed to the execution server and the execution programs will get updated automatically rather than reinstalling and recompiling the game apps. Therefore, most mobile apps support \textit{hotfix} for fixing code-related bugs. Particularly, for Unity games, we can insert bug code with the help of \textit{hotfix} without authority and knowledge of the source programs. We inject bug code by changing certain global variables or executing global functions from the Unity native interface. In this way, our code injection approach can be generalized to all Unity engine-based apps and even be easily adapted to graphics engines other than Unity by modifying the global variables and functions in the corresponding engines. Figure~\ref{fig:expaug} illustrate examples of augmented screenshots with UI glitch issues in which the first row shows the normal UI scenes and the second row displays the generated UI glitch scenes.

To be mentioned that GPU-related rendering bugs are typically bottom-layer issues or hardware setting problems and may vary differently from each game, also the manifestation of UI glitches caused by GPU-related rendering bugs is versatile depending on the device itself. Hence, we put this root cause in the future work, and our approach focuses on generating UI glitch scenes with the following three categories. 

\input{Others/figure9}

\definecolor{codegreen}{rgb}{0,0.6,0}
\definecolor{codegray}{rgb}{0.5,0.5,0.5}
\definecolor{codepurple}{rgb}{0.58,0,0.82}
\definecolor{backcolour}{rgb}{1,1,1}

\lstdefinestyle{mystyle}{
  backgroundcolor=\color{backcolour},   commentstyle=\color{codegreen},
  keywordstyle=\color{magenta},
  numberstyle=\tiny\color{codegray},
  stringstyle=\color{codepurple},
  basicstyle=\ttfamily\footnotesize,
  breakatwhitespace=false,         
  breaklines=true,                 
  captionpos=b,                    
  keepspaces=true,                 
  numbers=left,                    
  numbersep=5pt,                  
  showspaces=false,                
  showstringspaces=false,
  showtabs=false,                  
  tabsize=2
}

\lstdefinestyle{lua}{
  language=[5.1]Lua,
  basicstyle=\ttfamily,
  keywordstyle=\color{magenta},
  stringstyle=\color{blue},
  commentstyle=\color{black!50},
  backgroundcolor=\color{backcolour}
}

{\textbf{Turn off cameras in the scene.}} As we discussed before, the camera is used to capture the objects in the scene, we hence disable some cameras to force the UI glitches to appear in the game apps, we save the screenshots in different scenes as the abnormal samples. The patch code is shown in Listing 4. 

\lstset{style=lua}
\lstset{style=mystyle}
\begin{lstlisting}[xleftmargin=2.5ex, caption=Lua patch for turning off all cameras.]
local cameraobjs = CS.UnityEngine.Object.FindObjectsOfType(typeof(CS.UnityEngine.Camera))
for i = 0, cameraobjs.Length-1
do 
    cameraobjs[i].enabled=false
end
\end{lstlisting}

{\textbf{Modify camera's clearflag.}} Camera Clearflag, an \textit{enum} type, determines how to clear the depth buffer and color buffer before rendering the scene. There are four pre-defined values for clearflag: \textit{SkyBox}, \textit{SolidColor}, \textit{DepthOnly}, and Nothing. a) \textit{SkyBox}: clear the color buffer and depth buffer with \textit{SkyBox}; b) \textit{SolidColor}: clear the color buffer and depth buffer with \textit{SolidColor}; c) \textit{DepthOnly}: only clear the depth buffer; d) Nothing: don't clear either color buffer or depth buffer.
When we enter a scene, we traverse the camera, set the clearflag as one of the enum values we list above, and check the UI display state, if UI glitches occur, we save the screenshot as an abnormal example. The patch code is shown in Listing 5.

\lstset{style=mystyle}
\begin{lstlisting}[xleftmargin=2.5ex, caption=Lua patch for changing clearflag as DepthOnly for all existing camera in the scene.]
local  cameraobjs = CS.UnityEngine.Object.FindObjectsOfType(typeof(CS.UnityEngine.Camera))
for i = 0, cameraobjs.Length-1
do 
    cameraobjs[i].clearFlags=UnityEngine.CameraClearFlags.DepthOnly
end
\end{lstlisting}

{\textbf{Add incorrect post-processing effect.}} Adding post-processing effects is the last step in the Unity render pipeline, and the effects can modify the scene style easily. HDR and background blurring are two common post-processing effects. The UI image frame could become scrambled if we add incorrect post-processing effects to the scene (\eg adding the mirror effect to the button can lead to partial repetition). We randomly choose some post-processing effects and add them to different scenes and save the screenshot as the abnormal sample if the UI scene is messed up. The patch code is shown in Listing 6.
\lstset{style=mystyle}
\begin{lstlisting}[xleftmargin=2.5ex,caption=Lua patch example for adding mirror post-processing effect to current scene.]
local detectCamera = GameObject.Find("UICamera")
if (detectCamera ~= nil)
then
    detectCamera.gameObject: AddComponent(typeof(CameraFilterPack_3D_Mirror))
end
\end{lstlisting}

Note that there are two main reasons why we do not directly apply our summarized code patches to find bugs in game source code. First, injecting a type of bug requires to know only one code pattern of such bug type but detecting bugs requires knowing all patterns of this type of bug. In our patch, we only need to change some global variables or functions to generate UI glitches. However, for each bug type, there may be thousands of relevant statements and the correctness of each statement depends on other context codes. Second, even if one can figure out a code-analysis DL model for detecting bugs, the source code of the application under test is not always available (noted that our bug injection needs access to the source code of the training applications only). Thus, our GLIB is a more general and effective approach for real-world testing cases.

\subsection{CNN-Based UI Glitch Detection Model}
Deep Learning has achieved remarkable success in computer vision tasks such as image classification, object detection, object tracking, etc. and we hence choose the CNN architecture for detecting abnormal UI display images which can be regarded as one kind of image classification tasks. 

Given the screenshot as input to our CNN model, we firstly resize the input to a fixed size whose width and height is $w \times h$, then we use convolutional kernels to extract feature maps of the input followed by pooling layer which can progressively reduce the spatial size of feature representation meanwhile control overfitting. To improve the stability of CNN, the Batch Normalization (BN) is added after each convolutional layer. We obtain feature maps from the last convolutional layer and send them to the multiple full-connection (FC) layers to train a classifier with the $K$-dimensional vector as the output. Finally, the probability distribution of each class c is computed by softmax function:
 \begin{equation}
 P(y=c|x)=\frac{e^{f_c(x)}}{\sum_{k=1}^Ke^{f_k(x)}}\label{softmax}
 \end{equation}
 The classification result is given by the argmax function:
  \begin{equation}
 label = \mathop{\arg\max}_c  P(y=c|x).\label{argmax}
 \end{equation}
 To increase the nonlinearity of the CNN model, an activation function is added after the BN layer and FC layer.

\subsection{Saliency Map}
\label{subsec:saliency_map}
The CNN model only determines whether the image is abnormal, however, we are more concerned about which part of an image is abnormal thus can help the developer to fix the bug. Moreover, the saliency map~\cite{simonyan2013deep} can help to understand whether the model is accurate and how the model works. We simply compute the derivative of the label to the input by
   \begin{equation}
  \frac{\partial f(I)}{\partial I}\label{derivative}
 \end{equation}
 where $f$ represents the CNN model and $I$ is the input image. The bigger gradient value indicates a larger contribution to the classification result. If the output shows that the image is abnormal, the pixel with a large gradient value indicates the abnormal area.

\subsection{Implementation}

\input{Others/figure10}

We resize the input to a fixed size 512 × 256, if the image is vertical, we will rotate it to horizontal then resize it to the fixed size. Our CNN model consists of 10 convolutional layers whose kernel size is 3 × 3 followed by batch normalization layers, 5 MaxPooling layers, and 4 fully connected (FC) layers with the output size of the last layer $K = 2$, the activation function we adopt is ReLU. We set the number of kernel as 16 for convolutional layer $1\sim 4$, 32 for layer $5\sim 6$, 64 for layer $7\sim 8$ and 128 for layer $9\sim 10$. The model updates its parameter by minimizing the CrossEntropy loss for the two-label classification task. The network is trained by Adam optimizer over batches of 16 input images with an initial learning rate $\lambda$ as 0.001. The detail of our ConvNet configurations is shown in Figure~\ref{fig:net_arch_fig}. We implement our model based on the Pytorch framework.

%% file: Others/figure9.tex
\begin{figure}
     \centering
     \begin{subfigure}{0.155\textwidth}
         \centering
     \includegraphics[height=3.5cm, width=\textwidth]{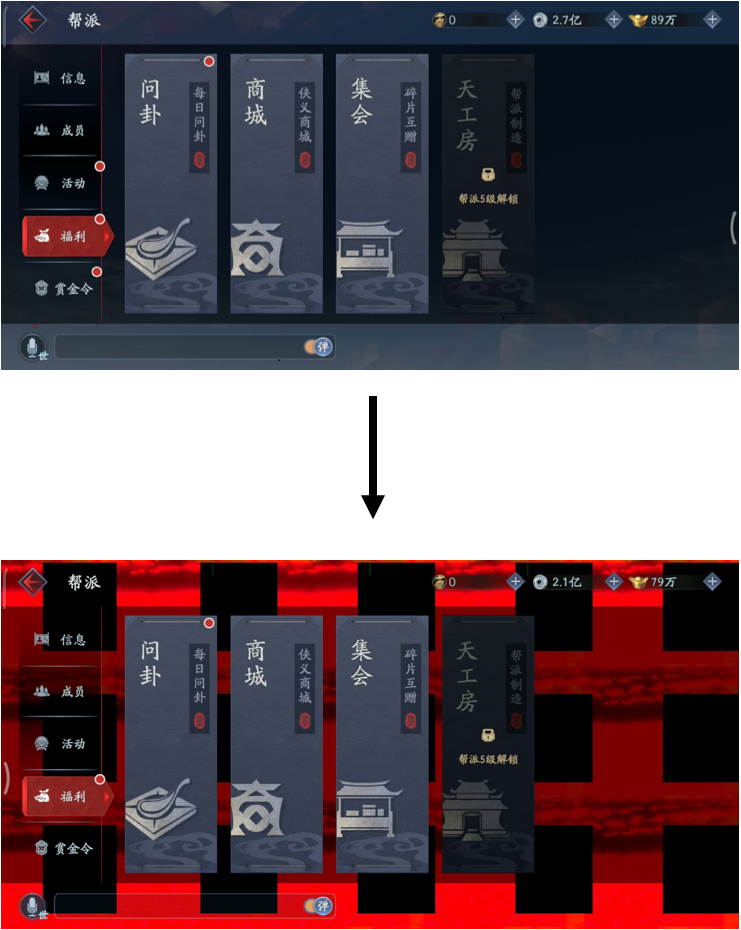}
         \caption{Camera enabled}
         \label{fig:cee}
     \end{subfigure}
     \hfill
     \begin{subfigure}{0.155\textwidth}
         \centering
     \includegraphics[height=3.5cm,width=\textwidth]{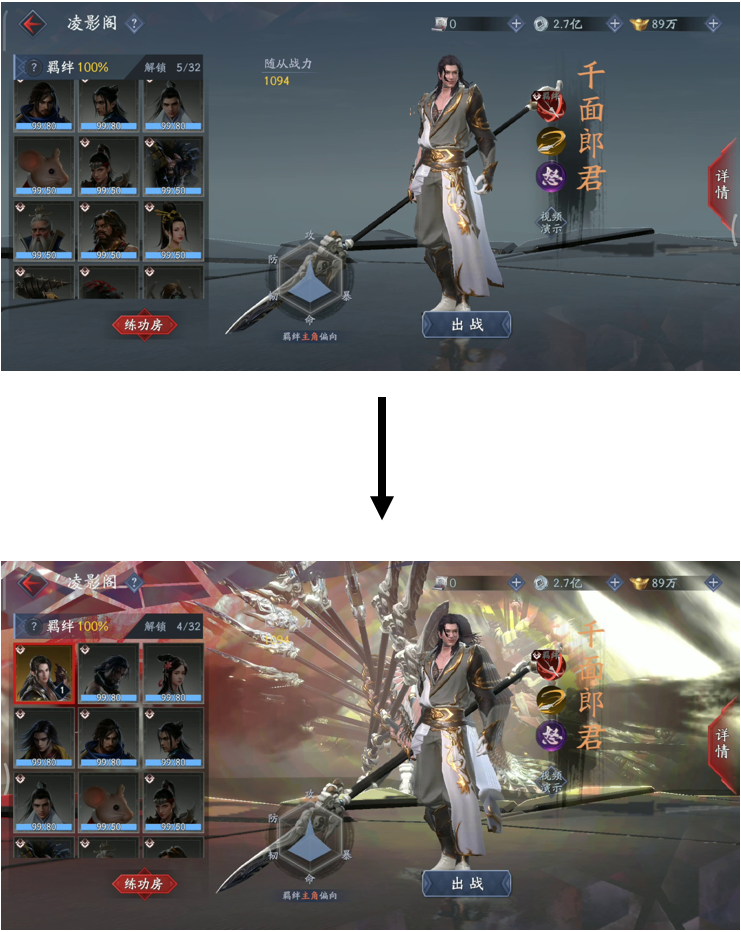}
         \caption{Camera clearflag}
         \label{fig:cce}
     \end{subfigure}
     \hfill
     \begin{subfigure}{0.155\textwidth}
         \centering
     \includegraphics[height=3.5cm,width=\textwidth]{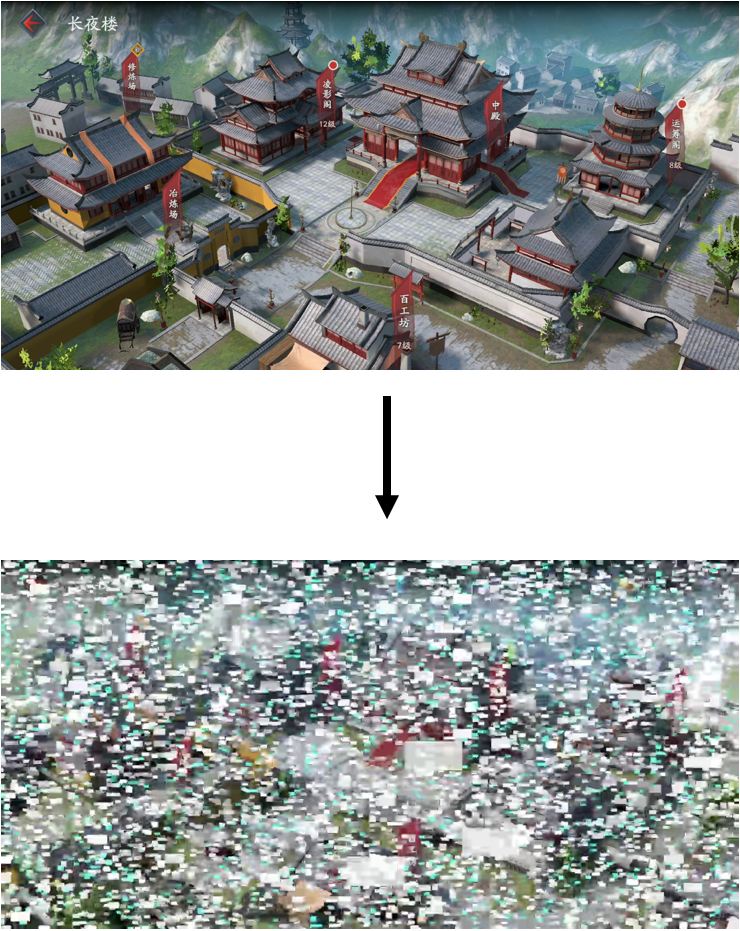}
         \caption{Post-processing}
         \label{fig:postprocess}
     \end{subfigure}
        \caption{Examples of code-based data augmentation.}
        \label{fig:expaug}
\end{figure}

%% file: Others/figure10.tex
\begin{figure}
     \centering
     \begin{subfigure}[b]{0.5\textwidth}
         \centering
     \includegraphics[width=\textwidth]{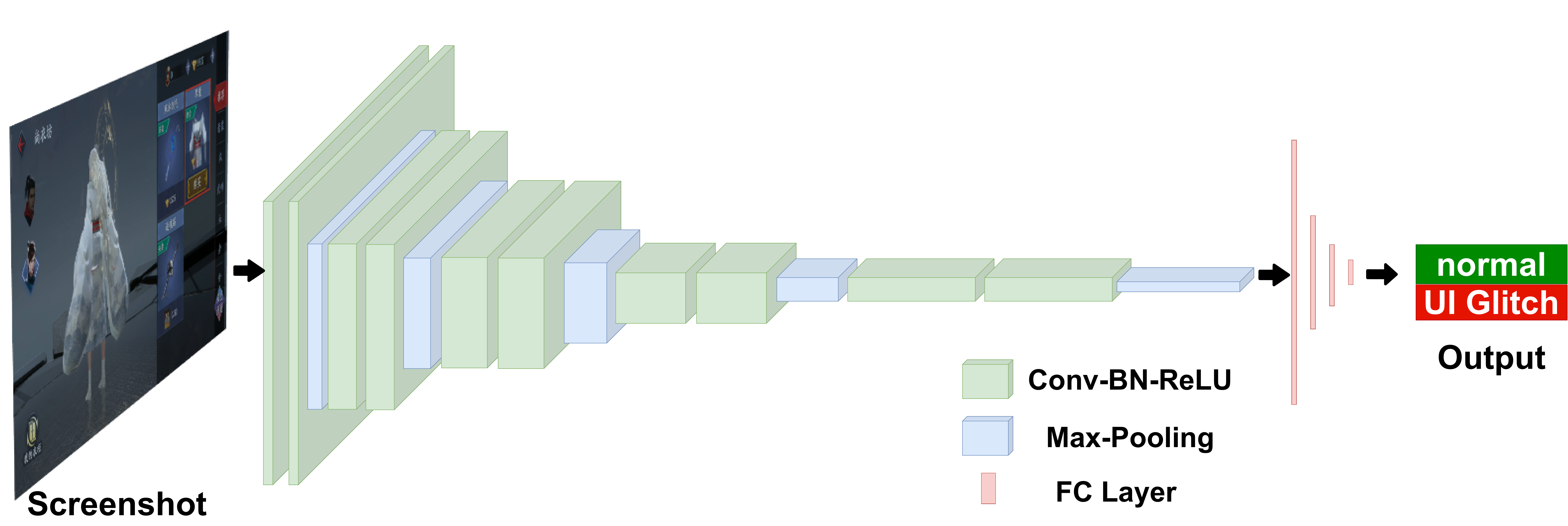}
     \end{subfigure}
        \caption{The architecture of CNN.}
        \label{fig:net_arch_fig}
\end{figure}

%% file: Evaluation.tex
\section{Experiment Design}

Our experiment is designed to answer the following questions:

\begin{itemize}[leftmargin=*]
    \item \textbf{RQ3:} How effective is \texttt{GLIB} in terms of detecting game UI glitches?
    
    \item\textbf{RQ4:} How well does \texttt{GLIB} perform in real-world game applications?
    
\end{itemize}

\subsection{Experimental Setup}
The game UI screenshots in our experiment are composed of 2 parts: screenshots without UI glitches (normal images) and screenshots generated by code-based augmentation approach (glitch images). We combine the two parts of data samples to train our \texttt{GLIB}. 

To balance the distribution of our training data, we roughly set the number of normal images and that of glitch images as 1: 1, particularly, we randomly select 6,841 screenshots from all the code-based generated glitch images for training, among which 2,511 of them are generated by setting the wrong clearflag of the scene camera, 3,763 of them are generated by turning the camera off and 567 of them are generated by adding incorrect post-processing effect. There are 6,817 normal screenshots and 6,817 glitch screenshots in the training dataset, 783 normal screenshots, and 759 glitch images in the validation dataset where 278 of them are generated by setting the incorrect camera clearflag, 418 are generated by turning the camera off and 63 are generated by adding incorrect post-processing effect.

\begin{table}
  \caption{Data Distribution}
  \label{tab:data_distribution}
  \begin{tabular}{cccc}
    \toprule
    Data Type & Augmentation Approach(s) & Game1 & Game2\\
    \midrule
    Normal &  & 1654 & 6186\\
    Glitch &  & 47 & 85 \\
    \midrule
        & Partial Repetition & 1654 & 6186\\
        & Solid Color Block & 1654 & 6186\\
Rule  & Mosaic Effect & 1654 & 6186\\
        & Random Noise & 1654 & 6186\\
    \midrule
        & Camera Turned Off & 1506 & 3056\\
Code  & Incorrect Camera Clearflag & 3144 & 1076\\
        & Incorrect Post-Processing & 330 & 300 \\
  \bottomrule
\end{tabular}
\end{table}

The screenshots are collected from two games, we manually traverse the scene in the mobile game app by clicking randomly on the mobile screen, capture the screenshots until UI components are stable, and save the screenshots as bug-free data. Then we apply three patches in Section~\ref{subsubsec:code_aug} to the game, if the screen is blurred, we capture the screen as the code augmentation result. Note that game scenes are typically dynamic rather than static. In each scene there may be multiple moving UI objects which produce a different screenshot in the next frame, thus there are quantities of different screenshots in each scene. Given that each game contains abundant different scenes, we can produce sufficient diverse abnormal screenshots for well-fitting the model. The rule-based data augmentation is an offline approach thus is processed after all the bug-free screenshots are collected. Table~\ref{tab:data_distribution} shows the distribution of screenshots we collected.

The test dataset that we use to evaluate the model is collected from 466 historical bug reports. We exclude the screenshots of game1 and game2 as well as some low-quality images and finally get 192 glitch images.

\begin{table}[H]
  \caption{Experiment Setup}
  \label{tab:experiment_data_setup}
  \begin{tabular}{cccc}
    \toprule
    Approach & Glitch Image & Normal Image & Total\\
    \midrule
    Base & 107 / 25 & 6817 / 300 & 6924 / 325\\
    Rule & 6817 / 783 & 6817 / 783 & 13634 / 1566\\
    Code & 6841 / 759 & 6817 / 783 & 13658 / 1542\\
    Code+Rule & 13658 / 1542 & 13634 / 1566 & 27292 / 3108\\
  \bottomrule
\end{tabular}
\end{table}

\subsection{Baselines}
To further demonstrate the advantage of our proposed data augmentation approach, we compare \texttt{GLIB} with five baselines utilizing deep learning techniques to examine the UI glitch detection effect. Because our goal is to detect UI glitches via bug understanding, for all the baseline we use the same CNN model and only with different data handling techniques. The dataset size of each baseline is listed in Table~\ref{tab:experiment_data_setup}.

\textbf{Base.}  We search the historical test reports of game1 as well as game2 and collect 132 screenshots which are truly bug images confirmed by the development teams. We select 125 of the 132 glitch screenshots and combine them with 6,817 normal images that we collect from game1 and game2 to build the training dataset without any augmentation approach, and our evaluation dataset consists of left 25 glitches screenshots and 300 normal images. We exclude the game1 and game2 screenshots from the 201 filtered graphical glitch screenshots which are collected from 20 game apps and remain 192 glitch screenshots for the test procedure.

\input{Others/figure11}

\textbf{Rule(R).} 
The heuristic-based data augmentation approach proposed by \texttt{Owl Eyes}~\cite{liu2020owl} contains several rules to approximate the UI display issues in non-game apps. Because most of their rules are based on text-relevant UI bugs (\eg NULL value, text overlap) which rarely appear in a game scenario, we adapt their rules to our studied manifestation of game UI glitches -- for each of the UI display issues, we generate screenshots by randomly choosing one of the following four rules. 
1) \textit{Image partial repetition}: we randomly choose a rectangle area in an image, then repeat sampling in a horizontal or vertical direction; 2) \textit{Adding solid color block}: we generate $3 \sim 5$ blocks where all pixels share the same color in one block and put them on an image randomly one by one, thus the former color block may be partial covered by the latter one. Color of each block can be arbitrary RGB value; 3) \textit{Adding mosaic effect}: we randomly choose a rectangle area in an image, dividing the area into several small patches where every pixel has the same RGB value as the center pixel in one patch; 4) \textit{Adding random noise}: we randomly choose a rectangle area in an image and set RGB value randomly for every pixel in the rectangle.
The four pairs of rule-based generated screenshots are shown in Figure~\ref{fig:dar}.

We generate abnormal samples based on the normal data with four simple heuristic approaches we discuss above, note that each normal screenshot is used only once. The final training data contains 6,817 normal screenshots and 6,817 generated glitch screenshots where 1,701 of them are generated by \textit{image partial repetition}, 1,659 of them are generated by \textit{adding solid color block}, 1,752 of them are generated by \textit{adding random noise} and the left 1705 of them are generated by \textit{adding mosaic effect} to the screenshots. The 1,566 evaluation samples are composed of 189 partial repetition glitch images, 184 abnormal color block screenshots, 220 random noise screenshots, 190 mosaic effect screenshots, and 783 normal screenshots.

\textbf{Rule(F).} For the second rule --\textit{adding solid color block}--we choose the RGB value of generated color blocks from the pre-defined four color values (red, black, pink and cyan) rather than arbitrary value due to the prior knowledge that these 4 color appears mostly when the material of game objects is missing or settled during the loading process. The other setting in this baseline is the same as Rule(R) approach.
 
\textbf{Code+Rule(R).} We combine both code-based and rule(R)-based generated screenshots as the training dataset for modeling the UI glitches. 

\textbf{Code+Rule(F).} We combine both code-based and rule(F)-based generated screenshots as the training dataset for modeling the UI glitches.

\subsection{Evaluation Metrics}
To evaluate the overall effectiveness of our proposed game UI display issue detection approach, we apply four commonly used evaluation metrics in image classification tasks, \ie accuracy, precision, recall, F1-score~\cite{ma2019easy,schutze2008introduction}. For all the metrics, a higher value indicates better model performance.

\textbf{Accuracy.} Accuracy reflects the trained model's ability to make correct decisions on the test set. The more correct samples the model predicts, the higher accuracy it will output.

\textbf{Precision.} Precision presents the proportion of correctly classified screenshots as UI glitch among all screenshots predicted as UI glitch.

\textbf{Recall.} Recall indicates the proportion of correctly classified screenshots as UI glitches among all screenshots that have UI display issues.

\textbf{F1-score (F-measure).} F1-score is calculated from the precision and recall of the test and it reflects the harmonic mean of precision and recall. The highest possible value of an F1-score is 1 which indicates perfect precision and recall, and the lowest possible value is 0 if either precision or recall is zero.

%% file: Others/figure11.tex
\begin{figure}
    \centering
    \includegraphics[width=0.475\textwidth,scale=0.25]{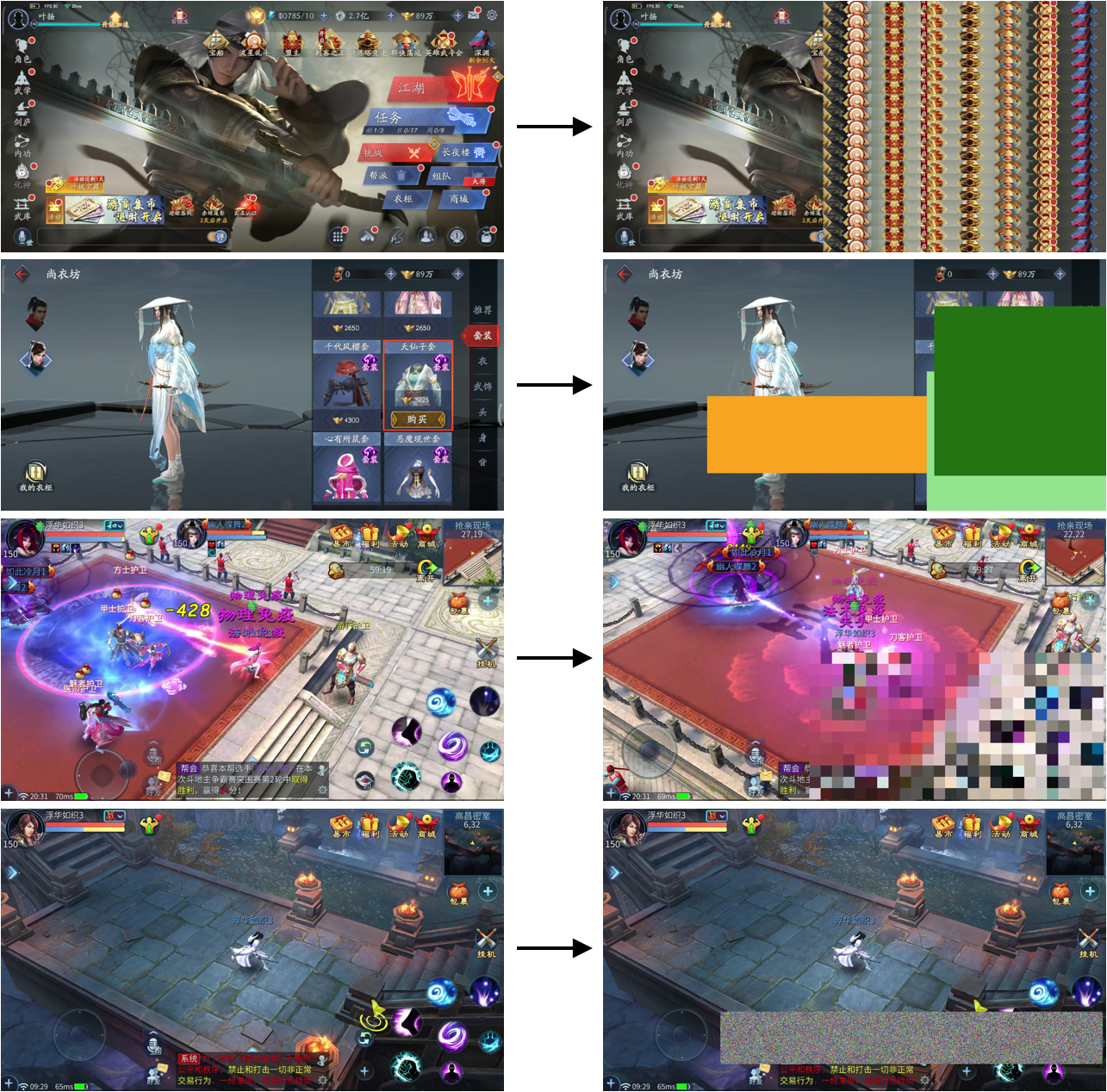}
    \caption{Examples of rule-based data augmentation, the four rows from top to bottom correspond to image partial repetition, adding solid color block, adding mosaic effect and adding random noise, respectively.}
    \label{fig:dar}
\end{figure}

%% file: Results_and_Analysis.tex
\section{Results and Analysis}
\subsection{UI Glitch Detection Performance (RQ3)}

We evaluate the effectiveness of our \texttt{GLIB} and the baseline approaches on the testing dataset composed of the collected 192 abnormal screenshots with UI glitches and 365 normal screenshots, the experiment results are listed in Table~\ref{tab:experiment_result}. 
We can see that our code-based data augmentation approach achieves the overall best performance (\ie highest precision/recall/F1\_score/accuracy). The 76.7\% increment of recall compared to the base approach indicate the effectiveness of our code-based data augmentation approach. We search for only one false negative sample and find that the error area in this screenshot is too tiny to be recognized even for humans.

Particularly, the five baseline approaches all achieve high precision, indicating that most screenshots predicted by the model as abnormal have UI glitches.
The base approach trained without any augmentation approach by only using the original glitch screenshots has the lowest recall (56.3\%), indicating that almost half of the UI glitch images are incorrectly classified by the model if not sufficient UI glitch samples are learned. We search the classification result and find that the model cannot detect the screenshots with UI glitches such as partial repetition, abnormal text, and abnormal color block. Though these categories occupy a large proportion of UI glitch issues, the number of glitch images is too small to cover the different patterns of these categories, and hence the model cannot learn the UI glitch manifestation sufficiently.

\begin{table}[h]
  \caption{Experiment Results}
  \label{tab:experiment_result}
  \begin{tabular}{ccccc}
    \toprule
    Approach & Precision & Recall  & F1\_score      & Accuracy\\
    \midrule
    Base & 0.991 & 0.563 & 0.718 & 0.803\\
    Rule(R) & 0.935 & 0.677 & 0.785 & 0.836\\
    Rule(F) & 0.974 & 0.594 & 0.738 & 0.813\\
    Code+Rule(R) & 0.990 & 0.984 & 0.987 & 0.988\\
    Code+Rule(F) & 0.995 & 0.948 & 0.971 & 0.975\\
    
    \textbf{GLIB} & \textbf{1.000} & \textbf{0.99} & \textbf{0.997} & \textbf{0.998}\\
  \bottomrule
  
\end{tabular}
\end{table}
 
Rule(F) incorrectly classifies 78 glitch screenshots as normal. We check these images and find that the glitch issues of them are mainly abnormal color block and text overlap. As we mentioned before, we only adopt four colors (black/red/pink/cyan) to generate the Solid Color Block in Rule(F) approach, which may cause the detection failure when a new color block appears. We check the bug reports and find that color blocks other than the pre-defined four types appear due to material missing. However, this unexpected error only occurs on a special GPU that has a special order of RGB values, which indicates that the manifestation of UI glitches on different devices caused by the same bug can still be different. The performance of Rule(F) is degraded by text overlap glitch issues because we do not consider the rules of abnormal text as we cannot localize the text area in UI screenshots without labeled JSON files that are typically not supplied by game engines. Moreover, in game apps some texts are displayed as Word-Arts or images but not the order of standard characters, thus the localization technique that uses OCR tools cannot work.

We study the 62 false-negative samples from Rule(R) and find that the main UI glitch categories of them are text overlap and abnormal color block, particularly, these color blocks are transparent and can be easily recognized as part of the background object. This transparent color block is similar to the dialog box in games and is misclassified also because we didn't generate the text-relevant glitch images in our training dataset. Because the transparent color blocks do not appear in the training dataset of the Rule approach, it is straightforward that the model cannot this type of UI glitches. Moreover, we find that the glitch images with blue color blocks are detected as UI glitch images whereas Rule(F) regards them as normal images, the reason may be that the Rule(R) approach can generate blue color blocks that can't be produced by Rule(F).

For Code+Rule(R) and Code+Rule(F), their recalls are largely improved compared to the single Rule(R) and Rule(F) approach, which demonstrates that the glitch images generated by our code-based approach can facilitate the model to learn more effectively. However, the reason that the combined approaches are not as effective as the single code-based augmentation approach may be that the distribution of code-based generated samples and rule-based generated samples are not identically consistent which may affect the training result.

\subsection{Practical Evaluation (RQ4)}
To examine the practical value of our \texttt{GLIB}, we collect two PC games from the official website, three iOS games from App Store~\cite{AppStore} and nine Android games from TapTap~\cite{TapTap} development teams. These games are developed by different game engines and none of these apps appear in the training dataset. 

Airtest~\cite{AirTest} is a cross-platform UI automatic game testing framework, and testers can write test scripts in Airtest IDE to execute specific test cases in the mobile device. Airtest IDE also provides a screen capture API for testers to take screenshots when necessary. We use the screen capture API to collect screenshots of various kinds of UI events (\eg click, swipe, long press, etc.) from the 14 games by running different test cases. We generate in total 2,100 screenshots from the 14 games, an average of 150 screenshots are obtained for each app. We then feed those screenshots to our \texttt{GLIB} for detecting abnormal UI issues. Once a UI glitch is spotted, we record the bug and report the issue to the app development team.

\begin{table}[H]
  \caption{Detected Game Issues}
  \label{tab:fixed or confirmed}
  \resizebox{\columnwidth}{!}{%
  \begin{tabular}{ccccc}
    \toprule
    Game Name & Game Category & Source & Daily Active Users & Download\\
    \midrule
    Justice & Role-Playing & Official Web & 300K+ & 50M+\\
    A Chinese Ghost Story & Role-Playing & Official Web & 300K+ & 50M+\\
    Ghost & Role-Playing & TapTap & 700K+ & 100M+\\
    Revelation Mobile & Role-Playing & TapTap & 500K+ & 10M+\\
    Love is Justice & Love & TapTap & 50K+ & 5M+\\
    UNO & Card & App Store & 50K+ & 5M+\\
    Fever Basketball & MOBA & App Store & 5K+ & 5M+\\
    Marvel Duel & Card & TapTap & 5K+ & 500K+\\
    Oracle Civilization & Simulation & App Store & 1k+ & 100k+\\
    Ghost World Chronicle & Card & Develop Team & N/A & N/A\\
    Elysium Of Legends & Card & Develop Team & N/A & N/A\\
    phase10 & Card & Develop Team & N/A & N/A\\
    Fpus & Shooting & Develop Team & N/A & N/A\\
    The Absolute Acting & Simulation & Develop Team & N/A & N/A \\
  \bottomrule
\end{tabular}
}
\end{table}

\begin{table}[h]
  \caption{Practical Evaluation Results}
  \label{tab:pratical_eval}
  \resizebox{\columnwidth}{!}{%
  \begin{tabular}{cccc}
    \toprule
    Platform & GLIB & Rule(R)  & Rule(F)\\
    \midrule
    PC & 7 & 3 & 1 \\
    Android & 35 & 22 & 15 \\
    iOS & 11 & 6 & 5 \\
    Total & 53 (48 confirmed) & 31 (28 confirmed) & 21 (17 confirmed) \\
  \bottomrule
\end{tabular}
}
\end{table}

Table~\ref{tab:fixed or confirmed} lists all bug issues detected by our \texttt{GLIB} and Table~\ref{tab:pratical_eval} shows the number of detected UI glitch issues on different platforms by the three approaches. In sum, \texttt{GLIB} detects 53 glitch issues and 48 of them are confirmed and fixed by the game development team; Rule(R) spots 31 glitch issues and 28 of them are confirmed and fixed; Rule(F) detects 21 glitch issues and 17 of them are confirmed and fixed. These confirmed and fixed bugs further demonstrate the effectiveness of the practical value of our proposed approach in detecting game UI glitches.

%% file: Case_study.tex
\section{Case Study}

To demonstrate that \texttt{GLIB} can accurately localize the glitch area of detected abnormal screenshots, we apply the saliency map introduced in Section~\ref{subsec:saliency_map} to show developers more detail about our model's prediction. We randomly select some images that are classified by \texttt{GLIB} as glitch images and calculate the derivative of the output concerning each pixel in the input image, the results are shown in Figure~\ref{fig:saliency_map_result} where the original screenshots are placed in the left and the generated saliency maps (which are converted to heat-maps) are listed in the right. A brighter area in the heat maps indicates a larger gradient of corresponding pixels, i.e, these pixels contribute more during the classify progress and are more likely to be the UI glitch issues.
The UI glitch of the first image in Figure~\ref{fig:saliency_map_result} is partial repetition, and the corresponding saliency map shows that \texttt{GLIB} concentrates much on these repetition areas, which is consistent with the manifestation. The saliency maps of the second and the third images indicate that the abnormal color blocks rather than the other background elements are the buggy area of the screenshots, which also agrees with the manifestation.
From the saliency map of the glitch screenshots, we can see that the model can not only accurately detect the image with UI glitch issues, but can also localize which part of the screenshot is abnormal.

\input{Others/figure12}

%% file: Others/figure12.tex
\begin{figure}[h]
     \centering
     \begin{subfigure}[b]{0.23\textwidth}
         \centering
     \includegraphics[width=\textwidth]{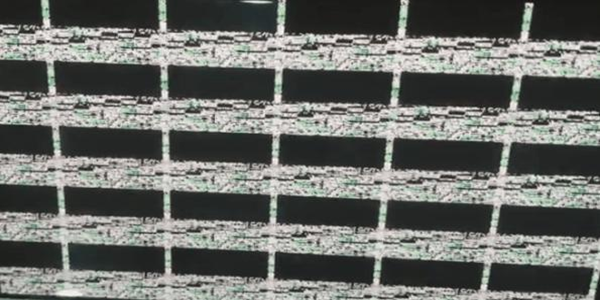}
         
         \label{fig:saliency_map_original3}
     \end{subfigure}
     \hfill
    \begin{subfigure}[b]{0.23\textwidth}
         \centering
     \includegraphics[width=\textwidth]{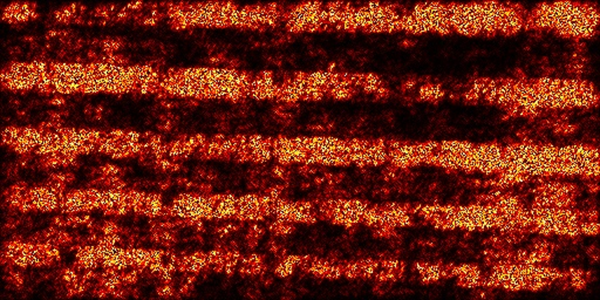}
         
         \label{fig:saliency_map3}
     \end{subfigure}
     \hfill
    \begin{subfigure}[b]{0.23\textwidth}
         \centering
     \includegraphics[width=\textwidth]{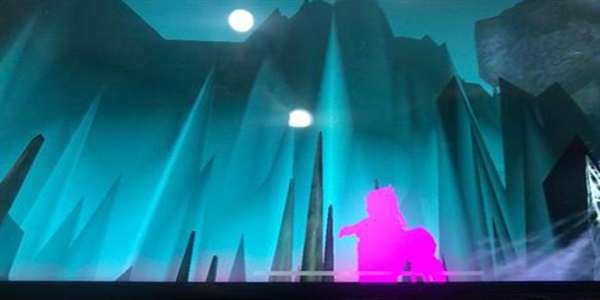}
         
         \label{fig:saliency_map_original2}
     \end{subfigure}
     \hfill
    \begin{subfigure}[b]{0.23\textwidth}
         \centering
     \includegraphics[width=\textwidth]{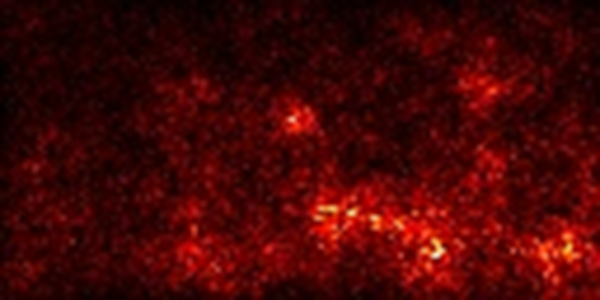}
         
         \label{fig:saliency_map2}
     \end{subfigure}
     \hfill
    \begin{subfigure}[b]{0.23\textwidth}
         \centering
     \includegraphics[width=\textwidth]{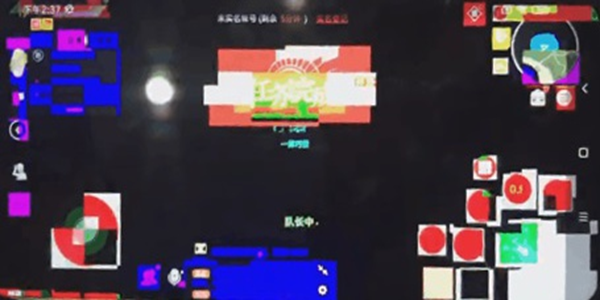}
         
         \label{fig:saliency_map_original4}
     \end{subfigure}
     \hfill
    \begin{subfigure}[b]{0.23\textwidth}
         \centering
     \includegraphics[width=\textwidth]{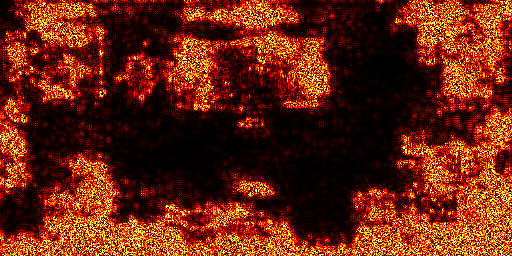}
         
         \label{fig:saliency_map_4}
     \end{subfigure}

        \caption{The images in the left column are the glitch images and images in the right column are the corresponding saliency map, the red pixel contribute most to the final label.}
        \label{fig:saliency_map_result}
\end{figure}

%% file: Application.tex
\section{Discussion}
In this section, we discuss the generality of our approach.

\textbf{Generality across games.} Our training data are collected and augmented from 2 Acting games, which may limit the model's applicability in real-world practice. However, our testing data used in RQ3 consists of 20 games including categories such as Role-playing games, card games, shooting games, MOBA (Multiplayer Online Battle Arena), love games, simulation games, etc. which nearly cover all popular daily-used game apps. The evaluation result shows that our proposed \texttt{GLIB} can accurately detect (99.5\% recall) all screenshots with UI glitches. This further demonstrates the generality of \texttt{GLIB} across different types of games. Moreover, our GLIB is a black-box image augmentation approach that requires source code only in the training phase and the well-trained model can be directly used to detect UI glitches on other games.

\textbf{Generality across languages.} Another advantage of our \texttt{GLIB} is that it can be applied for detecting UI glitches on game applications with different languages. Although the testing data of our experiment for RQ3 and case study only contains the screenshots of Chinese games, our study in Section~\ref{subsec:manifestation} shows that most of the game UI glitches are text-irrelevant, \ie abnormal text-only account for 2\% of all the UI display issues, plus our code-based data augmentation approach mainly focuses on UI glitches that are language-invariant such as abnormal color block, random noise, partial repetition and frame overlay caused by rendering effect or post-processing effect error. Hence, our proposed \texttt{GLIB} can be generalized for UI glitch detection in games with other languages.

\textbf{Generality across platforms.} 
Even though different games may run on different platforms, the game UI content is mostly decided by novel images which consist of 3d models as well as UI components. To prove that our model can precisely detect glitch images in terms of different game platforms, we collect UI screenshots from 9 Android games, 3 iOS games, and 2 PC games and use them as our testing dataset. The experiment results in RQ3 show that our approach can accurately detect all UI glitch images from games with different platforms, which further demonstrates the feasibility of our proposed \texttt{GLIB}.

\textbf{Generality across engines.} 
The game engine is a software development environment that provides developers with a series of tools to facilitate easy program writing. Games based on different engines may have different program structures, but the game UI rendering effect mainly depends on the low-level CPU/GPU system calls. For the same root cause (bug error), the manifestation of glitch images is typically similar regardless of which game engine the game is based on. To prove that our \texttt{GLIB} can well recognize glitch images across different engines, we select 2 games developed by a self-defined engine, 1 unreal-engine-based game, and 11 Unity games to compose our test dataset in RQ3. The experiment results show that our model can be generated well across different game engines.

%% file: Related.tex
\section{Related Work}

Our work, inspired by the automatic GUI testing~\cite{baek2016automated,mirzaei2016reducing,su2017guided} combined with deep learning technique, proposes a game GUI bug detection approach. GUI, a visual interface connecting users and software programs, has been studied by many researchers on different topics. Automatic GUI testing dynamically explores GUIs of an application, and several approaches~\cite{lamsa2017comparison,zein2016systematic} use computer vision techniques to detect GUI components to make predictions and compare different tools for GUI testing on Android applications. Recent deep learning-based techniques~\cite{degott2019learning,white2019improving} have also been applied for automatic GUI testing. More work on GUI with computer vision techniques such as GUI search~\cite{behrang2018guifetch,chen2019gallery,chen2020wireframe,chen2020lost,chen2020object,reiss2018seeking,zhao2019actionnet} and GUI code generation~\cite{chen2018ui,chen2019storydroid,chen2019gui,moran2018machine,nguyen2015reverse} facilitates the effective completion of computing tasks based on image features. 

On the other hand, many software linting tools aiming to flag bugs, stylistic errors, programming errors, and suspicious constructs~\cite{chen2020unblind,zhao2020seenomaly} have been proposed to ensure the normal operation of GUI. For example, StyleLint~\cite{StyleLint} helps developers avoid errors and enforce conventions in styles, Android Lint~\cite{Android_lint} reports over 260 different types of Android bugs including correctness, performance, security, usability, and accessibility. Different from static linting, our \texttt{GLIB} dynamically explores GUIs of an app as what automatic GUI testing does, but note that these GUI testing techniques concentrate on functional testing, whereas our work focuses on non-functional testing (\ie GUI glitches typically do not cause app crash but negatively affect the app usability). We analyze the GUI display issue in terms of software rendering bugs such as rendering camera settings and post-processing effects error which cause the app compatibility problems. It is extremely difficult and expensive for the developers to cover all the popular contexts when conducting testing. Moreover, our work only requires the screenshots as the input rather than these works based on static or dynamic code analysis. This crucial characteristic makes it easier for our lightweight CNN-based model to learn the pattern of UI glitch images and localize the UI glitches on the screenshot by a saliency map~\cite{simonyan2013deep} and also makes our approach more generalized to a different platform.

%% file: Threat.tex
\section{Threats to validity}

In our GLIB framework, the only manual part is to traverse and identify multiple diverse game scenes in each game for building our original training dataset. Also, our defined three categories of code injection approaches are based on study and empiricism. The code injection data augmentation based on \textit{hotfix} patching technique, CNN-based UI glitch detection as well as UI issue localization are all automated and can be easily adapted to other games on different platforms. Inappropriate selection of game screenshots in the manual part may weaken the \textit{external validity} of experimental conclusions. We try to mitigate this threat by traversing and selecting as many as distinct game scenes to make the dataset diverse and abundant. Our \textit{internal threat} mainly arises from the completeness of manifestation of game UI glitches and our defined three types of code patching approaches. We ask several game developing experts from NetEase to confirm that our summarized eight categories of UI glitches do cover all the common issues in their game apps. We also show that our code injection can generate all the most common five types of UI glitches.

%% file: Conclusion.tex
\section{Conclusion and Future Work}

Detecting and improving the quality of mobile game applications is of great value for nowadays game developers and testers. This paper proposes an automated UI glitch detection approach based on deep learning and bug analysis. Our empirical study on the root causes of game UI glitches facilitates a code-based data augmentation approach. Experimental results show that our \texttt{GLIB} is effective and shows great advantage in a real-world game testing scenario, \ie achieving nearly 100\% precision, recall, and F-1 score for classifying screenshots collected from 20 popular games, way better than the existing rule-based approach. Also, as the first work on game UI testing, we contribute to a systematical investigation of UI glitches in real-world mobile game apps, as well as a large-scale dataset of game app UIs with display issues for follow-up studies. Our proposed test oracle for automated UI glitch detection could facilitate further study on game UI testing.

In the future, we will focus more on the GPU-related rendering issues that cause game UI glitches and also keep improving the functionality of our model. Specifically, GLIB is one part of the game testing framework that we plan to research in the future. The whole framework contains an \textit{IO module}, a \textit{scene traverse module}, \texttt{GLIB}, and a \textit{log module}. First, \textit{IO module} captures screenshots from a mobile device and feeds them to both GLIB and the \textit{scene traverse module}; Second, GLIB classifies the given screenshot as normal or glitch (\textit{log module} loggings the corresponding information), and the \textit{scene traverse module} recognizes UI and click to yield the next scene, then it chooses a UI and returns the UI back to \textit{IO module}; Last, we repeat the two steps to realize the whole automated game testing.
Moreover, we hope to find a tight connection between bugs and the characteristic of UI glitches so that we can predict the bug code given a screenshot with UI display issues. And this bug inference technique can be more valuable when guiding developers to fix app compatibility issues.